\documentclass[a4paper,12pt]{article}

\usepackage{amsmath}\usepackage{amssymb}\usepackage[dvips]{graphicx}
\usepackage{latexsym}\usepackage{cite}\usepackage{bbold}

\usepackage{pstricks}
\usepackage{dcolumn}
\usepackage{bm}

\newcommand{\be}{\begin{equation}}
\newcommand{\ee}{\end{equation}}

\newcommand{\ka}{\kappa}

\newcommand{\mcF}{{\mathcal F}}
\newcommand{\mcH}{{\mathcal H}}
\newcommand{\mcK}{{\mathcal K}}
\newcommand{\mcL}{{\mathcal L}}
\newcommand{\mcN}{{\mathcal N}}
\newcommand{\mcO}{{\mathcal O}}
\newcommand{\mcV}{{\mathcal V}}
\newcommand{\mcW}{{\mathcal W}}

\newcommand{\half}{\frac{1}{2}}

\newcommand{\locsection}[1]{\setcounter{equation}{0}\section{#1}}

\def\beq{\begin{equation}}
\def\eeq{\end{equation}}

\def\Ga{\Gamma}

\def\ka{\kappa}
\def\si{\sigma}
\def\Si{\Sigma}
\def\te{\theta}
\def\La{\Lambda}
\def\lam{\lambda}

\def\l{\left (}
\def\r{\right )}

\def\fr{\frac}
\def\la{\label}
\def\hs{\hspace}
\def\vs{\vspace}

\def\ran{\rangle}
\def\lan{\langle}

\def\tl{\tilde}

\begin{document}
\begin{titlepage}
\begin{flushright}
HD-THEP-06-03\\
CERN-PH-TH-2006-024\\
February 17, 2006
\end{flushright}
\vspace{0.6cm}
\begin{center}
{\Large \bf 4D Superfield Reduction of 5D Orbifold SUGRA}\\ 
\vspace{12pt}{\Large \bf and}\\
\vspace{12pt}{\Large \bf Heterotic M-theory}
\end{center}
\vspace{0.5cm}

\begin{center}
{\large
Filipe Paccetti Correia$^a$\footnote{E-mail address:
paccetti@fc.up.pt},
Michael G. Schmidt$^b$\footnote{E-mail address:
m.g.schmidt@thphys.uni-heidelberg.de},
Zurab Tavartkiladze$^c$\footnote{E-mail address:
zurab.tavartkiladze@cern.ch}}

\vspace{0.3cm}

$^a${\em
Centro de F\' isica do Porto,
Faculdade de Ci\^ encias da Universidade do Porto\\
Rua do Campo Alegre 687, 4169-007 Porto, Portugal\\

$^b$
Institut f\"ur Theoretische Physik,
Universit\"at Heidelberg\\
Philosophenweg 16, 69120 Heidelberg, Germany\\

$^c$ Physics Department, Theory Division, CERN, CH-1211 Geneva 23, Switzerland
}
\end{center}
\vspace{0.4cm}
\begin{abstract}
We present a detailed study of the reduction to 4D of 5D supergravity compactified on the $S^1/{\mathbb Z}_2$ orbifold. For this purpose we develop and employ a recently proposed $\mcN=1$ conformal superfield description of the 5D supergravity couplings to abelian vector and hypermultiplets. In particular, we obtain a unique relation of the "radion" to chiral superfields as in global 5D SUSY and we can embed the universal hypermultiplet into this formalism. In our approach, it is transparent how the superconformal structure of the effective 4D actions is inherited from the one of the original 5D supergravity. We consider both ungauged and gauged 5D supergravities. This includes compactifications in unwarped geometries, generalizations of the supersymmetric Randall-Sundrum (RS) model as well as 5D heterotic M-theory. In the unwarped case, after obtaining the effective K\"ahler potentials and superpotentials, we demonstrate that the tree-level 4D potentials have flat and/or tachyonic directions. One-loop corrections to the K\"ahler potential and gaugino condensation are presented as suitable tools for moduli stabilization to be discussed in subsequent work. Turning to the RS-like models, we obtain a master formula for the K\"ahler potential for an arbitrary number of vector and hyper moduli, which we evaluate exactly for special cases. Finally, we formulate the superfield description of 5D heterotic M-theory and obtain its effective 4D description for the universal ($h^{(1,1)}=1$) case, in the presence of an arbitrary number of bulk 5-branes. We present, as a check of our expressions, time-dependent solutions of 4D heterotic M-theory, which uplift to 5D solutions generalizing the ones recently found in hep-th/0502077. 
\end{abstract}

%
%

\end{titlepage}

\tableofcontents

\setcounter{footnote}{0}

\locsection{Introduction}
The last decade has seen a great deal of interest in extra-dimensional set-ups with supersymmetry. Their study is not only justified by the fact that superstring/M-theory is consistent in more than 4D, but it is also backed by the fact that certain theoretical puzzles have a natural explanation within extra-dimensional constructions. This is specially true for braneworld models
(for string constructions see for example \cite{Horava:1995qa}), which have their own merits in explaning several issues. A list\footnote{An
appropriate list of references would be enormous and is not intended
here.} must include new aspects of supersymmetry breaking mediation \cite{Randall:1998uk,Gregoire:2004nn}, of the hierarchy problem \cite{Arkani-Hamed:1998rs,Randall:1999ee}, of Grand Unification \cite{Kawamura:2000ev}; the AdS/CFT correspondence \cite{Maldacena:1997re} should also be mentioned. Last but not least dynamical questions related to inflation \cite{Linde:2005dd}, the cosmological constant or quintessence gain a different face, as the effects of moving branes in cosmology \cite{Brax:2004xh} have to be studied. 5D constructions are a good laboratory for studying various effects of higher dimensions. 

Even if our final focus is towards model building, e.g. for inflation \cite{us05a}, 5D supersymmetry has to be embedded in 5D supergravity in order to have a firm basis. In our line of research \cite{us04a,us04b} we have developed a formalism with 4D superfields. This approach was obtained by studying the off-shell component 5D supergravity of refs.\cite{Kugo:2000hn,Kugo:2000af,Fujita01} (for other pioneering work in this direction see also \cite{Zucker:1999ej,Zucker:1999fn,Zucker:2000ks}), and their reduction of the multiplets of 5D conformal supergravity to 4D multiplets of the $\mcN=1$ conformal supergravity surviving at the boundaries of the $S^1/{\mathbb Z}_2$ orbifold \cite{Kugo:2002js}. For the purposes of the present work, it is necessary to use an equivalent but slightly modified version of the formulation of \cite{us04a}. This is obtained by integrating out a multiplier real superfield ${\mathbb W}_y$ - called real \emph{radion} superfield in \cite{us04a} - to get a superspace action where the radion field $e^5_y$ is now contained only in \emph{chiral} superfields. In this way we clarify the meaning of the superfield ${\mathbb W}_y$ and make clear that a trully \emph{dynamical} radion field is already included in the formulation of \cite{us04a}.  

However, 5D supersymmetry or supergravity models are more than just well-meant string theory inspired models displaying nice phenomenological features. In fact, heterotic M-theory compactified on a Calabi-Yau 3-fold has a well defined low-energy description in terms of a 5D gauged supergravity on the $S^1/{\mathbb Z}_2$ orbifold \cite{Lukas:1998yy,Lukas:1998tt}. Also, it is expected \cite{Brummer:2005sh} that the physics of (strongly) warped throats obtained by compactifying the type IIB string on certain CY 3-folds with fluxes be described by a \emph{gauged} 5D supergravity with a non-trivial scalar manifold \cite{Ceresole:2000jd}, at low energies. 

Generally such theories with extended SUGRA contain numerous moduli, parametrizing the volume, shape and other properties of the compactification manifold. The formulation of a dynamical stabilization mechanism leading to a stable ground state is of utmost importance - in the simplest case in 5D the stabilization of the radion mentioned above, but more general also its interplay with other moduli (e.g. in gauge inflation \cite{us05a}).      

Going to lower energies, below a certain "compactification" scale $M_c$ one enters the reign of 4D physics. In this work, we are interested in the case that below $M_c$ physics can be described by an effective 4D $\mcN=1$ supergravity theory. Knowing this theory, in priciple one should be able to understand supersymmetry breaking and its transmission to the visible sector, and to calculate all relevant couplings to compare with experiments. Eventually, also inflationary (and post-inflationary) cosmology can be studied within the effective 4D theory. It is therefore imperative to have \emph{a clear understanding of the connection between the 5D supergravity theory and its 4D daughter supergravity theory}. 

Building on our recent work \cite{us04a} on a \emph{conformal} superfield description of 5D supergravity theories, in this paper we study mostly the dimensional reduction of these theories, obtaining in this way the corresponding effective 4D \emph{superconformal} supergravities. An obvious advantage of this approach, is that since we use the same $\mcN=1$ superfield language both in 5D and in 4D, the connection between the 5D action and its 4D counterpart becomes \emph{transparent}. This is true, in particular, for the way the 4D superconformal structure emerges from its higher-dimensional origin. 

Even if not exhaustive, our investigation is rather general, comprising generic abelian vector and hyper scalar manifolds as well as different types of R-symmetry gaugings. It encompasses, therefore, both unwarped and warped geometries, which we will study in different sections. The particularly interesting case of a warped geometry, provided by 5D heterotic M-theory, will deserve a separate investigation in section 5. 

The basic priciple behind the concept of dimensional reduction of a 5D (or higher-dimensional) theory is the assumption that at sufficiently low energies the theory is well described by the interactions of a reduced (finite) set of 4D fields. Our task consists of of indentifying these fields and their interactions. In the case at hand, 5D supergravity, we have to identify low-energy superfields, in particular the holomorphic moduli (chiral superfields) and, of course, the supersymmetry of the interactions must be ensured. Usually, the main difficulty lies in going beyond the \emph {static} approximation, where the 4D action is obtained by solving the 5D equations of motion in the background of constant 4D moduli, and inserting the solutions into the 5D Lagrangian which is then integrated over the compact $S^1/{\mathbb Z}_2$ orbifold. Promoting the static moduli to "slowly" varying functions of the 4D coordinates generally leads to non-vanishing VEVs of certain heavy fields $C_{\mu}\sim f\partial_\mu \phi$. This is especially true for warped geometries, as in this case $f$ can depend on the warped wave-functions of the zero-modes, and the VEVs cannot be simply gauged away. The way these tadpoles backreact into the low-energy physics, leading to new two-derivative order terms, is e.g. explained in an enlightening discussion in \cite{Luty:2000ec} (see also \cite{Bagger00}). This is an important technical problem in any compactification with substantial warping (see \cite{Giddings:2005ff} for a recent discussion in the context of the type IIB warped compactifications of \cite{Giddings:2001yu}).

We will see in this paper how the task of reducing 5D supergravity models on the $S^1/{\mathbb Z}_2$ orbifold can be immensely simplified by employing the $\mcN=1$ conformal superfield formulation of the 5D supergravity couplings to (abelian) vector and hypermultiplets of \cite{us04a}. In fact, the low-energy supermultiplets, in particular the holomorphic moduli chiral superfields, can be easily identified, in virtue of starting already in 5D with $\mcN=1$ superfields. In addition, as we argue in section \ref{sec:warp}, the "tadpole-problem" discussed above can be bypassed by noting that, at the level of a two derivative truncation, the K\"ahler potential can be determined using just the static approximation. In very special cases, namely for the supersymmetric Randall-Sundrum model and the universal case of heterotic M-theory, it turns out that we do not even need an explicit knowledge of the moduli wave-functions to obtain the corresponding effective K\"ahler potentials in an exact manner.

In this way, we are e.g. able to determine the K\"ahler potential of heterotic M-theory in the universal case with an arbitrary number of bulk 5-branes, which has the following suggestive form
\be\nonumber
                      \mcK=-3\ln\left(\sum_i\beta_i\left(S^i+S^{i+}\right)^{\frac{4}{3}}\right),
\ee
where $V^i=\textup{Re}(S^i)$ is the volume of the Calabi-Yau 3-fold measured at the $i$-th brane (including the boundary branes). We use this result to obtain \emph{new} time-dependent solutions of 4D heterotic M-theory. They uplift to 5D solutions generalizing the ones recently found in \cite{Chen:2005jp}. 

While it is possible to obtain exact expressions for the K\"ahler and superpotential for 5D orbifold SUGRA with an arbitrary number of vector and hypermultiplets in case there is \emph{no} warping, in the warped case this is only possible in a handfull of cases. In this paper we give general formulae for the K\"ahler potential in warped compactifications. We think these will be usefull as a starting point for future approximate approaches to this problem. This is particulary important if one wants to obtain explicitely the K\"ahler potential for more general compactifications of 11D heterotic M-theory, beyond the linear approximation of \cite{Lukas:1997fg}.

\vs{12pt}

The paper is organized as follows: In section 2 we review, improve and extend our superfield description of 5D supergravity coupled to vector and hypermultiplets. It is also explained with a specific example how to take the rigid limit within this formalism. We then move on to perform the dimensional reduction of 5D supergravity on the $S^1/{\mathbb Z}_2$ orbifold for geometries without warping in section 3. In addition to tree-level K\"ahler potentials and superpotentials, we obtain 1-loop corrections to the K\"ahler potential and discuss gaugino condensation. Section 4 is devoted to the dimensional reduction of 5D supergravities with warped geometries of RS type. Finally, in section 5 we present a superfield formulation of 5D heterotic M-theory and derive its reduction to 4D in the universal case with arbitrary number of bulk 5-branes. This enables us to obtain new time-dependent solutions of 4D heterotic M-theory and their uplift to the 5D theory. Appendix A contains a proof of the generic flatness or instability of tree-level 5D supergravity without warping.

\locsection{5D orbifold (conformal) supergravity}\label{sec:unwarp}

For the applications that we have in mind in this paper it is necessary to recall 5D ($\mcN=2$) SUGRA in its description in terms of 4D $\mcN=1$ superfields, as it was given in \cite{us04a}. But we will also go beyond that work to obtain new \emph{expressions}, which despite being equivalent to \cite{us04a} turn out to be better suited for the dimensional reduction to 4D that we will perform in section \ref{sec:reduction}. Also new is the inclusion of the universal hypermultiplet in this formalism. This will allow us to give a superfield formulation of 5D heterotic M-theory and of the corresponding 4D effective theory in section \ref{sec:mtheory}. We close this section with a derivation (and improvement) of well-known superspace actions for 5D SYM theories coupled to charged hypermultiplets and the radion multiplet, obtained by performing the rigid limit of 5D orbifold supergravity. 

\subsection{Our framework}
We start this section by recalling the couplings of abelian vector multiplets and hypermultiplets in 5D (conformal) supergravity in the superfield description introduced in \cite{us04a}. We will consider first the case with no \emph{physical} hypermultiplets, i.e. in addition to the vector multiplets we take only \emph{one}\footnote{We will have to consider \emph{two} compensator hypermultiplets (in addition to one physical hypermultiplet) to obtain the universal multiplet (see sec.\ref{sec:universal}).} compensator hypermultiplet, which - as usual in the superconformal formalism - will be used to fix the conformal symmetries. This will be explained in some detail below. We also assume that there are no gauged $R$-symmetries, leaving the more general situation to subsequent sections.

The vector part of the Lagrangian reads,
\be\begin{split}\label{eq:lagrange_vec}
                 \mcL_V=&\frac{1}{4}\int d^2\theta \,\left(-\mcN_{IJ}(\Sigma)\mcW^{\alpha I}\mcW_{\alpha}^J+\frac{1}{12}\mcN_{IJK}{\bar D}^2(V^ID^{\alpha}\partial_yV^J-D^{\alpha}V^I\partial_yV^J)\mcW^K_{\alpha}\right)\\
                        &\qquad+\textup{h.c.} -\int d^4\theta \,{\mathbb W}_y\,\mcN ({\mcV}_5),
\end{split}\ee
while for the compensator hypermultiplet we have
\be\label{eq:lagrange_comp}
                    \mcL_{comp}= -2\int d^4\theta \,{{\mathbb W}}_y\,({ h}^+ { h}+h^{c+} h^c)-2\left(\int d^2\theta \,h^c\partial_y h+\textup{h.c.}\right).
\ee
These as well as all other superspace integrals in this paper should be understood as F and D-densities of 4D (conformal) SUGRA in the sense of refs.\cite{Ferrara:1977ij,Kaku:1978nz,Kaku:1978ea,Cremmer:1978hn,Kugo:1982mr,Kallosh:2000ve}. 

A few words are due to explain the meaning of these expressions. Let us first remember that the indices $I,J$, etc..., run over the different abelian\footnote{This formalism can be easily extended to include also non-abelian vector multiplets.} vector multiplets ${\mathbb V}^I$ of off-shell 5D supergravity. It is important to note that as the \emph{graviphoton} is entailed in (a combination of) the vector multiplets, their number is allways greater than one. The ${\mathbb V}^I$ decompose in vector superfields $V^I$, and chiral superfields $\Sigma^I$ of $\mcN=1$ (conformal) supersymmetry: ${\mathbb V}^I=(V^I,\Sigma^I)$. (For the detailed constitution of each superfield see \cite{us04a}.) The other two basic building blocks of the above Lagrangians are the chiral superfields $h$ and $h^c$ which together form the 5D compensator hypermultiplet ${\mathbb h}=(h,h^c)$, and the real \emph{radion} superfield ${\mathbb W}_y$. The reason we call ${\mathbb W}_y$ a radion superfield is that its first component is $e^5_y$:
\be\label{eq:oldradion}
        {\mathbb W}_y= e^{-\sigma}e^5_y +\theta\,2\ka_5e^{-\frac{\sigma}{2}}\psi^1_y+{\bar\theta}\,2\ka_5e^{-\frac{\sigma}{2}}{\bar\psi}^1_y+\cdots   
\ee 
Here, $e^{2\sigma}$ accounts for a possible warping\footnote{Note that the superfields used in this paper are written in terms of the component fields of 5D SUGRA in such a way that no explicit warp-factors appear in \eqref{eq:lagrange_vec} and \eqref{eq:lagrange_comp}. It turns out that, as one can see in \eqref{eq:oldradion}, each field comes with a power $\exp(w\,\sigma)$ of the warp-factor, where $w$ is the Weyl-weight of the corresponding field. In fact, \eqref{eq:lagrange_vec} and \eqref{eq:lagrange_comp} were first obtained for unwarped geometries. To use these Lagrangians also for warped geometries requires using a Weyl transformation to bring the warped metric into a flat form, as explained in detail in appendix D in ref.\cite{us04a}. The superfields in this section thus are the primed ones of that appendix.}, 
\be\label{eq:metric5D}
                ds_{5}^2=e^{2\sigma}\,g_{\mu\nu}dx^{\mu}dx^{\nu}+(e^5_ y)^2\,dy^2,
\ee
$\psi^1_y$ and ${\bar\psi}^1_y$ are obtained from the 5D gravitino and the dots stand for a number of gravitational auxiliary fields, see \cite{us04a}. $\ka_5^2=M_5^{-3}$ is the five-dimensional gravitational coupling. It is important to stress $e^5_y$ enters the Lagrangian not only through ${\mathbb W}_y$ but also through the chiral superfields $\Sigma^I$. In fact we have 
\be\label{eq:def_sigma}
          \Sigma^I=\half (e^5_y M^I+iA_y^I)+\cdots,
\ee
where $M^I$ is the scalar component of the 5D vector multiplet, $A_y^I$ the $y$-component of the gauge boson. After conformal fixing in the sense of ref.\cite{Fujita01}, a cubic combination of the $M^I$ ($I=0,\dots,n_V$) will be fixed. 

The remaining superfields in the above Lagrangians are composite ones: $\mcW^{\alpha I}$ is the usual chiral superfield strength obtained from the vector superfield $V^I$, while 
\be
         {\mcV}_5^I=\frac{\Sigma^I+\Sigma^{I+}-\partial_yV^I}{{\mathbb W}_y}.                
\ee

The strength and form of the couplings are set by the so-called \emph{norm} function $\mcN$. This is a homogeneous gauge invariant cubic function of its arguments
\be 
          \mcN(\phi)=\ka_5\, c_{IJK}\phi^I\phi^J\phi^K,
\ee     
where $\phi^I$ can have different meanings: $\phi^I=M^I,\Sigma^I,{\mcV}_5^I$. We will see below that it is intimatelly related to the prepotential of global $\mcN=2$ supersymmetry, which is the reason sometimes it is called prepotential. As we pointed out above, after conformal fixing, the vector scalars $M^I$ will be constrained. This constraint is given by
\be\label{eq:constr}
            \mcN(M)=\ka_5^{-2},
\ee 
which defines a very special manifold (an $n_V$-dimensional hypersurface in the space parameterized by the scalars $M^I$). We will comment on this issue again below. We use the notation $\mcN_I=\partial_I\mcN(M)$ etc., to denote the derivatives of the norm function. As pointed out in \cite{us04a} the combination $\Sigma_{rad}=(\mcN_I/3\ka_5\mcN)\Sigma^I$ approaches the \emph{usual} radion chiral superfield in the rigid susy limit, but this is not a central issue in our approach, we just deal with the superfields in the Lagrangian above.\\

The above Lagrangians, as they stand, proved to be quite usefull for studying various issues in 5D, see \cite{us04a,Abe:2004ar,us04b,us05a,Abe:2005wn}. As we will now explain - and this is a new point - there is an equivalent form of these Lagrangians which for certain purposes (e.g. to obtain the 4D effective actions) turns out to be more usefull than \eqref{eq:lagrange_vec} and \eqref{eq:lagrange_comp}. It is obtained by \emph{integrating out} the real superfield ${\mathbb W}_y$. In fact, a variation of the action with respect to this superfield implies the following
superfield identity\footnote{Note that this identity is true also if we take the couplings to the supergravity multiplet into account, i.e. there are no tree-level modifications due to supergravity. However, this relation is modified once we include 1-loop corrections, as we will see in section \ref{subsec:kaehlerstuff}.}:
\be\label{eq:wy}
           {\mathbb W}_y=\left(\frac{\mcN(V_y)}{h^+h+h^{c+}h^c}\right)^{\frac{1}{3}},
\ee
where we introduced $V_y^I\equiv \Sigma^I+\Sigma^{I+}-\partial_yV^I={\mathbb W}_y{\mcV}_5^I$. Inserting this back in the D-term Lagrangians we get:
\be\label{eq:Dlagran_1}
              \mcL_{D}=-3\int d^4\theta \,\mcN(V_y)^{\frac{1}{3}}\left[h^+h+h^{c+}h^c\right]^{\frac{2}{3}}.
\ee
This can easily be extended to include also physical hypermultiplets ${\mathbb H}=(H,H^c)$,
\be\label{eq:Dlagran_hypers}
               \mcL_{D}=-3\int d^4\theta \,\mcN(V_y)^{\frac{1}{3}}\left[h^+h+h^{c+}h^c-H^+e^{-g_IV^I}H-H^{c+}e^{g_IV^I}H^c\right]^{\frac{2}{3}},
\ee 
where we assumed the hypermultiplet to be charged under some combination of the abelian gauge symmetries.
The main advantage of this \emph{new} form of the Lagrangian is that now all the information on the radion is stored in the chiral superfields $\Sigma^I$. In this way in the following sections we will - in principle - be able to obtain the low energy 4D effective K\" ahler potentials from general 5D field contents. 

Before we end this section, let us for completeness explain how one fixes the superconformal symmetries to obtain the 5D (Poincar\' e) supergravity theory. Let us however note that in the rest of the paper we will perform the reduction of the theory to 4D \emph{prior} to superconformal gauge fixing. The following lines are therefore usefull if one wants to stick to 5D. As already pointed out, the fixing of the superconformal symmetries is achieved by prescribing certain VEVs to the fields in the hypermultiplet \cite{Fujita01}. To understand how to do this we first recall that (as we already mentioned) all the superspace integrals in this paper should be understood \cite{Kugo:1982mr} as F- and D-densities of 4D conformal supergravity. In particular, for a real superfield $Q=\phi_Q+\cdots +\te^2{\bar\theta}^2 D_Q$ we have
\be
             \int d^4\te\, Q\to(Q)_D=\int d^4\te \,Q+\fr{1}{3}\phi_QR^{(4)}+\cdots,
\ee
where $R^{(4)}$ is the Ricci scalar build from the 4D metric $g_{\mu\nu}(x)$ in \eqref{eq:metric5D}. Taking $Q$ as the integrand superfield in eq.\eqref{eq:Dlagran_hypers}, we see that, to ensure that we are in the 5D Einstein-frame, one should set\footnote{The 5D Ricci scalar is related to 4D by $R^{(5)}=e^{-2\sigma}R^{(4)}+\cdots$, where the dots stand for terms involving derivatives of the warp-factor. These additional terms emerge naturally from our superspace Lagrangians, see \cite{us04a,us04b}.} 
\be
             \mcN(M)^{\frac{1}{3}}\left[|\phi_h|^2+|\phi_h^c|^2-|\phi_H|^2-|\phi_H^c|^2\right]^{\frac{2}{3}}=\ka_5^{-2}\,e^{2\sigma},
\ee
where $\phi_h,\phi_H$ etc., are the scalar components of the chiral superfields $h$ and $H$, respectively. The prescription of \cite{Fujita01} used in \cite{us04a} is to take
\be\label{eq:fixing_1}
                 h=\ka_5^{-1}\,\left[e^{3\sigma}+\ka_5^2(|\phi_H|^2+|\phi_H^c|^2)\right]^{\half}+\theta^2\,F_h,\qquad h^c=\theta^2\,F_h^c,
\ee  
along with the constraint of eq.\eqref{eq:constr}. It is important here to emphasize the difference between eqs.\eqref{eq:fixing_1} and \eqref{eq:constr}: From the point of view of the superspace formalism, eq.\eqref{eq:fixing_1} is truly a superconformal fixing, while \eqref{eq:constr} is just a definition. Namely, denoting by $\half\sigma^I$ the real part of the scalar component of $\Sigma^I$ (i.e. $\Sigma^I=\half\sigma^I+\cdots$), then we can define $\left(e^{5}_y\right)^3 \equiv\ka_5^2\mcN(\sigma)=\ka_5^3\,c_{IJK}\sigma^I\sigma^J\sigma^K$ and the constraint \eqref{eq:constr} follows from this definition. The superfields $\Sigma^I$, however, remain unconstrained. In fact, the splitting of the $\sigma^I$ in $e^5_y$ and $M^I$ is necessary just to make the connection with the conventional formulations of 5D supergravity in component form \cite{Fujita01}, where it plays a fundamental r\^ole.

\subsection{The universal hypermultiplet}\label{sec:universal}

The so-called universal hypermultiplet is an important ingredient of string and M-theory compactifications to 5D supergravity. In particular, in the compactification of 11D supergravity on a Calabi-Yau 3-fold, one of its scalar components describes the volume of this manifold. In contrast to the "hypers" of eq.\eqref{eq:Dlagran_hypers} which parametrise the coset USp(2,2)$/$USp(2)$\times$USp(2), the universal hyperscalar spans an SU(2,1)$/$U(2) manifold. It was pointed out in ref.\cite{Fujita01} that to obtain such a hyperscalar manifold one has to introduce a second compensator hypermultiplet. We will denote the two compensators by ${\mathbb h}_i=(h_i,h_i^c)\sim(+,-)$, $i=1,2$. In addition, to fix the additional unphysical degrees of freedom one introduces a multiplier vector multiplet ${\mathbb V}_T=(V_T,\Sigma_T)\sim(+,-)$, which does not participate in the norm function $\mcN(M)$. All hypermultiplets will be charged under the ${\mathbb V}_T$ with charges $q({\mathbb H})=q({\mathbb h}_1)=-q({\mathbb h}_2)$, and the corresponding D-term Lagrangian will read
\be\begin{split}\label{eq:univer_Dprim}
          \mcL_{D}= & -3\int d^4\theta \,\mcN(V_y)^{\frac{1}{3}} \left[h_1^+e^{-V_T}h_1+h_1^{c+}e^{V_T}h_1^c+h_2^+e^{V_T}h_2\right.\\
	            &\hspace{96pt}\left.+h_2^{c+}e^{-V_T}h_2^c-H^+e^{-V_T}H-H^{c+}e^{V_T}H^c\right]^{\frac{2}{3}},
\end{split}\ee   
while the F-term Lagrangian is given as
\be\label{eq:univer_Fprim}
          \mcL_{F}=-2\int d^2\theta \,\left\{ h_1^c\partial_y h_1 + h_2^c\partial_y h_2 -H^c\partial_y H-\Sigma_T(h_1^c h_1-h_2^c h_2-H^c H) \right\}+\textup{h.c.}
\ee
As we already said, $V_T$ and $\Sigma_T$ are not dynamical, they only couple to the hypers, and it is thus possible to integrate them out exactly. Their superfield equations of motion imply
\be\label{eq:multiplier}
                e^{2V_T}=\frac{h_1^+h_1-H^+H+h_2^{c+}h_2^c}{h_1^{c+}h_1^c-H^{c+}H^c+h_2^+h_2},
\ee 
and
\be
              h_1^c h_1-h_2^c h_2-H^c H=0.  
\ee
Before we insert this back in the Lagrangians let us redefine the chiral superfields in the hyper sector with
\be
              h=(2h_1h_2)^{\frac{1}{2}}, \quad h^c=\frac{h_2^c}{h_1}, 
\ee
for the compensators, and
\be
             \Phi=\frac{H}{h_1},\quad  \Phi^c=\frac{H^c}{h_2},  
\ee
for the physical superfields. Due to the above mentioned constraints the Lagrangians can be rewriten in terms of this \emph{reduced} set of superfields. In fact, we have now
\be\label{eq:univer_D}
             \mcL_{D}=-3\int d^4\theta \,\mcN(V_y)^{\frac{1}{3}} (h^+h)^{\frac{2}{3}} \left\{\left(1-|\Phi|^2+|h^c|^2\right)\left(1-|\Phi^c|^2+|h^c+\Phi\Phi^c|^2\right) \right\}^{\frac{1}{3}},
\ee
and
\be\label{eq:univer_F}
             \mcL_{F}=-2\int d^2\theta \,\left( hh^c\partial_y h-\frac{h^2}{2}\Phi^c \partial_y\Phi \right)+\textup{h.c.}
\ee
Note that for notational simplicity we used $H^+H\to|H|^2$ and similar for other \emph{superfields}. Finally, to obtain the component Lagrangian in terms of the "traditional" complex scalars $S$ and $\xi$, we must use the following \emph{non-holomorphic} variable redefinitions \cite{Fujita01}
\be
                 (\Phi)_{\theta=0}=\frac{1-S}{1+S},\qquad (\Phi^c)_{\theta=0}=\frac{\xi(1+S)}{S+S^+},       
\ee
supplemented by the conformal gauge fixing conditions (similar to \eqref{eq:fixing_1}) 
\be
                (h)_{\theta=0}=\ka_5^{-1}\left\{\left(1-|\Phi|^2\right)\left(1-|\Phi^c|^2+|\Phi\Phi^c|^2\right)\right\}^{-\frac{1}{4}}, \qquad (h^c)_{\theta=0}=0.
\ee

\subsection{The rigid limit in 5D}

We will see now how well-known results regarding the superfield description of 5D super Yang-Mills theories \cite{Arkani-Hamed:2001tb,Marti:2001iw,Hebecker:2001ke,Dudas:2004ni} as well as their coupling to the radion superfield \cite{Marti:2001iw} can be obtained in the \emph{rigid} limit of 5D supergravity with a particular choice of the norm function and of the orbifold parities of the involved superfields. To take the rigid limit of 5D supergravity we choose a supersymmetric "vacuum" and then perform an expansion of the Lagrangian in powers of $\ka_5=M_5^{-3/2}$. To be precise, consider $\ka_5^{-1}\mcN(M)=(M^0)^3-M^0\delta_{ij}M^iM^j + \gamma_{ijk} M^iM^jM^k$, $i,j,k\neq 0$, and take the parity prescriptions $T\equiv 2\ka_5\Sigma^0\sim +$ and $\Sigma^i\sim -$. This implies that $\gamma_{ijk}=\epsilon(y)c_{ijk}$ is an odd coupling. In addition we include one charged hypermultiplet as in eq.\eqref{eq:Dlagran_hypers} with $g_0=0$ and $g_i\neq 0$. These models generally have several supersymmetric vacua with $H=H^c=0$, which must satisfy (see e.g. \cite{us04b})
\be
                 \partial_y \mcN_I(M)=0, \textup{ with } I=0,i,
\ee
under the constraint $\mcN(M)=\ka_5^{-2}$. We will expand here around the $M^i=0$ vacuum, i.e. we assume that $M^i\ll M^0$ which in turn implies $T+T^+\gg 2\ka_5(\Sigma^i+\Sigma^{i+})$. For the term $\mcN(V_y)^{\frac{1}{3}}$ in eq.\eqref{eq:Dlagran_hypers} this means
\be
            (\ka_5^{-1}\mcN(V_y))^{\frac{1}{3}}=\frac{T+T^+}{2\ka_5}-\frac{2\ka_5}{3}\delta_{ij}\frac{V^i_y\,V^j_y}{T+T^+}+\frac{4\ka_5^2\epsilon}{3}c_{ijk}\frac{V^i_y\,V^j_y\,V^k_y}{(T+T^+)^2}+ \cdots, 
\ee
where, we recall, $V_y^i=\Sigma^i+\Sigma^{i+}-\partial_yV^i$ is a gauge invariant superfield \cite{Hebecker:2001ke}. In addition, since there is no $R$-symmetry gauging ($g_0=0$) and therefore also no warping, we can set the compensator superfields to $h=\ka_5^{-1}$ and $h^c=0$, and expand the hypermultiplet part of \eqref{eq:Dlagran_hypers} in powers of $H^+H$ and $H^{c+}H^c$, obtaining in this way
\be
        \left[\ka_5^{-2}-H^+e^{-g_iV^i}H-H^{c+}e^{g_iV^i}H^c\right]^{\frac{2}{3}}=\ka_5^{-\frac{4}{3}}-\frac{2\ka_5^{\frac{2}{3}}}{3}\left(H^+e^{-g_iV^i}H+H^{c+}e^{g_iV^i}H^c\right)+\cdots
\ee
We can now assemble these pieces to obtain the leading contributions to the D-term Lagrangian
\be\begin{split}
                   \mcL_D\simeq\int d^4\theta \,(T+T^+)&\left[-\frac{3}{2}M_5^3+2\delta_{ij}\frac{V^i_y\,V^j_y}{(T+T^+)^2}-4\epsilon(y)\ka_5 \,c_{ijk}\frac{V^i_y\,V^j_y\,V^k_y}{(T+T^+)^3}\right.\\ &\hspace{144pt}\left.+H^+e^{-g_iV^i}H+H^{c+}e^{g_iV^i}H^c\right]+\cdots
\end{split}\ee 
Clearly, while we dropped higher order terms which are not given in \cite{Arkani-Hamed:2001tb,Marti:2001iw,Hebecker:2001ke,Dudas:2004ni}, these are fully computable within our formalism. Note that a first derivation of this expression was given in \cite{us04a}, even though in a less elegant manner. It is not difficult to see that the vector part of the Lagrangian can be rewritten in terms of a prepotential (see also \cite{PaccettiLobodeMendoncaCorreia:2005sv})
\be\label{eq:prepo}
                      \mcF(\phi)=3M_5^3-\delta_{ij}\phi^i\phi^j+\epsilon(y)\ka_5\,c_{ijk}\phi^i\phi^j\phi^k,
\ee
as
\be
                  \mcL_V = -\frac{1}{4}\int d^2\theta\left(T\mcF_{ij}\left(\Sigma/T\right)\mcW^i\mcW^j+\frac{1}{12}\mcF_{ijk}(\cdots)\right)+\textup{h.c.}-\int d^4\theta \frac{T+T^+}{2}\mcF\left(\frac{2V_y}{T+T^+}\right),
\ee
where $\mcF_{ij}=\frac{\partial^2\mcF}{\partial\phi_i\partial\phi_j}$, etc. The first term in the r.h.s. of eq.\eqref{eq:prepo} corresponds to the pure radion part, first given in \cite{luty99}, the second term to the radion interaction with vector multiplets \cite{Marti:2001iw}. The third is new and describes the Chern-Simons-type interaction, involving also the radion.

\locsection{Reduction to 4D in unwarped geometries: K\"ahler and superpotentials}\label{sec:reduction}
In the rest of this paper we will show how to obtain the effective 4D supergravity description of 5D orbifold supergravity models within the above formalism. As we will see, the superfield approach turns out to be well suited for this purpose, specially in the unwarped case.

We first set the notation we will use throughout this and the following sections. The 4D Lagrangian is determined by three functions of the moduli, namely the K\"ahler potential, the superpotential and the gauge kinetic function. The \emph{K\" ahler} potential $\mcK(\Phi,\Phi^+,V)$ is defined by
\be\label{eq:eleven}
          \mcL_{D}^{(4D)}=- 3\int d^4\theta \,e^{-\mcK/3}\,\phi^+\phi,     
\ee
where $\phi$ is the 4D chiral compensator, while the gauge kinetic function $f_{IJ}(\Phi)$ and the superpotential $W(\Phi)$ are defined through
\be
          \mcL_F^{(4D)}=\frac{1}{4}\int d^2\theta \,f_{IJ}{\mcW}^{\alpha I}{\mcW}_{\alpha}^J+\int d^2\theta\phi^3\, W + \textup{h.c.}        
\ee

Since in this section we consider unwarped geometries, the wave-functions of the zero modes of the various superfields (even under ${\mathbb Z}_2$-orbifolding) will be $y$-independent. In the following we consider a dimensionless $e^5_y$ and therefore we have to introduce a \emph{fixed} length $R$ so that $-\pi R<y<\pi R$. We emphasize that $\pi R$ is not dynamical, but for practical reasons we will take it to coincide with the size of the extra-dimension (after this is stabilized). All odd superfields will be projected out\footnote{We assume the VEVs of the odd superfields to vanish. If this is not the case we can redefine the superfield to ensure that it does so.}. In particular, the odd compensator $h^c$ will be set to zero. Denoting by 
\be
        \phi=\sqrt{2\pi R}\ka_5^{-\frac{1}{3}}h^{\frac{2}{3}},
\ee	
the 4D compensator and by ${\tilde\Sigma}^I=\ka_5\Sigma^I$ the 4D vector moduli, we get from \eqref{eq:Dlagran_1}
\be
             \mcL_{D}^{(4D)}=-3\int d^4\theta \,[{\tilde\mcN}({\tilde\Sigma}+{\tilde\Sigma}^+)]^{\frac{1}{3}}\phi^+\phi,
\ee
where ${\tilde\mcN}$ is obtained from $\ka_5^{-1}\mcN$ by setting the odd fields to zero. In the following we will drop the tilde in $\tilde\Sigma$, hoping that the reader doesn't get confused by this. In the simplest case with $n_V=0$ and $\mcN=\ka_5 (M^0)^3$ we get
\be
           \mcL_{D}^{(4D)}=-3\int d^4\theta \,(\Sigma^0+\Sigma^{0+})\phi^+\phi,
\ee
a well known expression \cite{luty99}. Note that in this case $\Sigma^0$ is the \emph{dynamical} (4D) radion superfield.

In this way, in the models discussed above, the effective tree-level K\" ahler potential is
\be\label{eq:kaehler_simplest}
                 \mcK(\Sigma,\Sigma^+)= -\ln{\tilde\mcN}(\Sigma+\Sigma^+), 
\ee
agreeing with known results (see \cite{brax04} and references therein). It is straightforward to generalize this K\"ahler potential to include
also physical hypermultiplets. Then the whole (tree level) 
K\"ahler potential is
\beq
{\cal K}=-\ln \tl{\cal N}(\Si +\Si^+)-
2\ln \l 1-\Phi^{+}e^{-g_IV^I}\Phi+\cdots\r ~,
\la{genK}
\eeq
where $\Phi=H/h$, and we assumed $H^c$ to be the odd superfield. 

Similarly, denoting the 4D field-strength superfield by ${\mcW_{4D}}=\sqrt{2\pi R}\mcW$, we obtain also the following gauge kinetic term
\be
          f_{IJ}(\Sigma)=-{\hat\mcN}_{IJ}(\Sigma),         
\ee
where $\hat\mcN_{IJ}(\Sigma)$ is obtained from $\ka_5^{-1}\mcN(\Sigma)$ by differentiating twice in respect to odd $\Sigma^I$ and then setting the odd superfields to zero. (Note that in general ${\hat\mcN}_{IJ}(\Sigma)\neq{\tilde\mcN}_{IJ}(\Sigma)$.) 

By gauging certain isometries of the scalar manifold it is also possible to obtain tree-level 4D superpotentials and 4D Fayet-Iliopoulos terms directly from the 5D theory. These can then lead to non-vanishing tree-level potentials. We will discuss these issues in the following subsection.

\subsection{Tree-level potentials and (in)stabilities}\label{sec:towards}

We will now write down the 4D \emph{tree-level} potentials obtained from the 5D bulk supergravity without (physical) hypermultiplets upon compactification on the $S^1/{\mathbb Z}_2$ orbifold. From a 5D point of view there are essentialy \emph{three} kinds of models, namely Minkowski (${\mathbb M}_4$), de Sitter (dS) and anti de Sitter (AdS). Both the dS and AdS cases turn out to be unstable non-supersymmetric compactifications, obtained by certain R-\emph{symmetry} gaugings to be precised below. In the no-scale (supersymmetric) ${\mathbb M}_4$ vacua there are no gauged R-symmetries. Depending on whether the gauged U(1)$_R$ is broken by orbifolding or not, the generated potential will be an F-term or a D-term potential. Notice that if it was not for the orbifolding, due to the (in that case unbroken) SU(2)$_R$ symmetry the potentials would be the same in both cases. The action of the orbifold twist has subtle consequences and therefore we will study the two cases separately.

\subsubsection{Superpotentials and F-term potentials}
 
Let us recall that the full scalar potential of 4D supergravity consists of two parts: an F-term and a D-term potential. We will work here in the 4D Einstein frame. This is obtained by fixing the lowest component of the chiral compensator to be $\phi=M_P+\cdots$. In this frame, the F-term potential is given in terms of the K\"ahler potential $\mcK$ and the superpotential $W$ as
\be\label{eq:formula}
                 V_F=M_P^4\,e^{\mcK}\left(\mcK^{I{\bar J}}D_IWD_{\bar J}{\bar W} - 3|W|^2\right), \quad \textup{with } D_I\equiv\partial_I+\mcK_I,       
\ee
where $I$ runs over both hyper and vector scalars ($I=H,\Sigma$). We will discuss D-term potentials later in section \ref{sub:FIterms}.

Our investigation will focus on the \emph{vector} moduli $\Sigma$. One reason for this is that the \emph{hyper} moduli $H$ can be stabilized in a fairly simple way by using tree-level \emph{brane} superpotentials. In respect to this, one should remember that the vector moduli are less simple to couple directly to the branes (see however section \ref{sec:susystab}).

In fact, due to the no-scale nature of the $\Sigma$, which means that at tree-level
\be
               \mcK^{\Sigma{\bar\Sigma}}\mcK_{\Sigma}\mcK_{\bar\Sigma}=3,
\ee 
an $H$-dependent superpotential $W(H)$, which has to be brane localized, will generate a \emph{positive} definite potential for the hyper moduli, since
\be
                 V_F=M_P^4\,e^{\mcK}\mcK^{H{\bar H}}D_HWD_{\bar H}{\bar W},        
\ee
while the D-term potential is positive by definition, see \eqref{eq:vidi}. \emph{Supersymmetric} vacua with the hypermultiplets $H$ fixed at suitable values can be obtained by fine-tuning $W(H)$. At the end of this procedure, the vector moduli remain flat directions which still must be stabilized. The effects studied in this and later sections, namely $\Sigma$-dependent superpotentials, 1-loop corrections to the K\"ahler potential and gaugino condensation, might be usefull to achieve this purpose.

For simplicity, let us start with a field content consisting of vector multiplets ${\mathbb V}^I=(V^I,\Sigma^I)\sim(-,+)$, where $+$($-$) denote the ${\mathbb Z}_2$-orbifold \emph{parities} of the corresponding superfields, and the compensator hypermultiplet $(h,h^c)\sim(+,-)$. In this case, the relevant bulk interaction is given by
\be
                        \mcL=-3\int d^4\theta\,\mcN(V_y)^{\frac{1}{3}}\left[h^+h+h^{c+}h^c+\cdots\right]^{\frac{2}{3}}-\left(\int d^2\theta \,(2h^c\partial_y h-g_I\Sigma^I (h^2-h^{c2}))+\textup{h.c.}\right),
\ee
where the last 2 couplings on the r.h.s. are due to gauging by the combination $g_I\mathbb{V}^I$ of an U(1)$_R$ subgroup of the SU(2)$_R$ in the $\sigma^{1,2}$ (Pauli matrices) direction. This U(1)$_R$ is projected out by the action of the orbifold. Clearly, by setting $g_I=0$ we recover the ungauged case. At this point it is important to note that by coupling a certain $\mathbb{V}^{I=R}$ to a linear multiplet $\mathbb{L}$ it is possible to impose the constraints 
$V^R=0$ and $g_R\Sigma^R=i\omega/R$ (i.e. only lowest component of $\Sigma^R$
is non zero), where $\omega$ is a real quantity which can be viewed as a \emph{Scherk-Schwarz} parameter \cite{Scherk78}. A detailed explanation of this mechanism is given in \cite{us05a}, where the necessary ingredients for radion mediated supersymmetry breaking - equivalent to SS breaking - within 5D orbifold SUGRA are given. 

The effective 4D couplings are then easily shown to be given by the K\" ahler potential $\mcK=-\ln{\tilde\mcN}(\Sigma+\Sigma^+)$ (see eq.\eqref{eq:kaehler_simplest}) and the superpotential $W$, defined by
\be\label{eq:def_superpotent}
                       W={\bar g}_I \Sigma^I+i\ka_4\frac{\omega}{R},
\ee
where the 4D gauge coupling is given by ${\bar g}^{-2}=2\pi R g^{-2}$. The corresponding potential, obtained from eq.\eqref{eq:formula} with a few manipulations, is\footnote{In the derivation we used that $\mcK^{I{\bar J}}=-{\tilde\mcN}\,{\tilde\mcN}^{I{\bar J}}+\frac{1}{2}(\Sigma+\Sigma^+)^I(\Sigma+\Sigma^+)^{\bar J}$.} 
\be\begin{split}\label{eq:FtermPot1}
               V_F & =M_P^4\,e^{\mcK/M_P^2}\left({\bar g}_I\mcK^{I{\bar J}}{\bar g}_{\bar J}-({\bar g}_I(\Sigma^I+\Sigma^{I+}))^2\right)\\
	       & =-M_P^4{\bar g}_I{\tilde\mcN}^{I{\bar J}}{\bar g}_{\bar J}-\frac{M_P^4}{2\tilde\mcN}({\bar g}_I(\Sigma^I+\Sigma^{I+}))^2,
\end{split}\ee
where, let us recall, $\tilde\mcN=c_{IJK}(\Sigma^I+\Sigma^{I+})(\Sigma^J+\Sigma^{J+})(\Sigma^K+\Sigma^{K+})$. Notice that this (tree-level) potential doesn't depend on the Scherk-Schwarz parameter $\omega$, even for ${\bar g}_I\neq 0$.

As we stated above, models with ${\bar g}_I=0$ are of the no-scale type \cite{Cremmer:1983bf,Ellis:1983sf,Dragon:1984vx,Dragon:1986ew,Lahanas:1986uc}, and therefore have a tree-level vanishing potential. On the other hand, turning on the gauge couplings ${\bar g}_I$ one can achieve non-zero potentials. However, in this case one finds instabilities. Before we proceed with the general analysis it would be instructive to illustrate this with two simple \emph{examples} where the above potential can be determined explicitly:
\begin{itemize}
\item[(A)] The first one is ${\tilde\mcN}=(\sigma_0)^3$, where $\sigma_I\equiv2\textup{Re}(\Sigma^I)$, and we get
\be\label{eq:exampleA}
                 V_{(A)}=-\frac{2}{3}\frac{M_P^4{\bar g}_0^2}{\sigma_0}.  
\ee
\item[(B)] In the second case we take ${\tilde\mcN}=(\sigma_0)^3-\sigma_0(\sigma_1)^2$, obtaining for ${\bar g}_0=0$
\be
                 V_{(B)}=\frac{M_P^4 {\bar g}_1^2}{2\sigma_0}\left[\frac{1}{1+\tfrac{1}{3}\psi^2}-\frac{\psi^2}{1-\psi^2}\right]\simeq\frac{M_P^4 {\bar g}_1^2}{2\sigma_0}\left(1-\tfrac{4}{3}\psi^2+\cdots\right),
\ee
where $\psi\equiv \sigma_1/\sigma_0<1$. Another convenient parameterization of the scalar manifold is
\be
\si_0=\rho\cosh^{2/3}t~,~~~~~~\si_1=\rho\fr{\sinh t}{\cosh^{1/3}t}~,
\la{parManif}
\ee
where $\rho$ accounts for the radion $e_y^5$. This parameterization  satisfies the constraint $\tl{\cal N}=\rho^3$. The exact potential becomes now
\be
V_{(B)}=\fr{M_P^4\bar g^2}{2\rho}\fr{3-4\sinh^4t}
{\cosh^{2/3}t\,(3+4\sinh^2t)}~.
\la{V2xt}
\ee
\end{itemize}

Clearly, $V_{(A)}$ and $V_{(B)}$ exhibit instabilities which in both cases will lead to the collapse of the extra-dimension: $\sigma_0\to 0$ in the first case, $\rho\to 0$ ($t\to\infty$) in the second one. Back to 5D, $V_{(A)}$ corresponds to a bulk negative cosmological constant, while $V_{(B)}$ describes a bulk positive cosmological constant and a tachyonic scalar (to be parametrized by $\psi$). As we will show now, this is a generic feature when, for at least some $I$, ${\bar g}_I\neq 0$. 

To be more precise, we will prove here that if $\Sigma_0^I$ is at an \emph{extremum} of the potential $V_F$, i.e. $\frac{\partial V_F}{\partial\Sigma_0^I}=0$, there is at least one \emph{tachyonic} direction at $\Sigma_0^I$. This means that it is not possible to simultaneously stabilise all the vector moduli. (Below we will see that the addition of a constant superpotential can remedy this situation, leading to AdS vacua.) A detailed derivation of this assertion is provided in appendix \ref{app:instability}, while here we only present the crucial steps. Essentially, the argument goes by showing that at any extremum the potential $V_F$ must vanish:
\be
                       V_F(\Sigma_0)=0, \textup{ for } \frac{\partial V_F}{\partial\Sigma_0^L}=0.
\ee
Using this fact we then go on to find that for the \emph{vector} $v^{J}\equiv {\bar g}_I{\tilde\mcN}^{I}(\Sigma_0+\Sigma_0^+)$ we have
\be
                      v^J\frac{\partial^2 V_F}{\partial\Sigma_0^J\partial{\bar\Sigma}_0^{\bar I}}v^{I}\leq 0,
\ee
where the bound is saturated only in the no-scale case ($g_I=0$ $\forall_I$). We see that the mass$^2$-matrix at the extremum $\Sigma_0$ displays at least one negative eigenvalue (unless we consider the no-scale case).

This situation changes by inclusion of a moduli independent
brane superpotential $W_{br}$ with non vanishing real part. For example,
\begin{itemize} 
\item[(C)] with $\tl{\cal N}=(\si_0)^3$ and adding to (A) the superpotential
$W_{br}$, the potential becomes
\beq
V_{(C)}=-M_P^4\l \fr{2}{3}\fr{\bar g_0^2}{\si_0}+
\fr{\bar g_0}{\si_0^2}(W_{br}+W_{br}^+)\r ~.
\la{C}
\eeq
\end{itemize}
For $g_0(W_{br}+W_{br}^+)<0$ the radion $\si_0$ is stabilized at
$\lan \si_0\ran =-3(W_{br}+W_{br}^+)/\bar g_0$. However, this vacuum is
AdS. 
For a self consistent uplifting to a ${\mathbb M}_4$ or dS vacuum, additional care is needed.

\subsubsection{FI terms and D-term potentials}\label{sub:FIterms}

We consider now vector multiplets ${\mathbb V}^M=(V^M,\Sigma^M)\sim(+,-)$, which differ from the ones in the previous paragraphs in their orbifold parities. Besides that, we still have vector multiplets of the previous type ${\mathbb V}^I=(V^I,\Sigma^I)\sim(-,+)$, which entail the radion and other moduli. We have the bulk interactions:
\be\begin{split}\label{eq:FIterms1}
               \mcL= & -\int d^4\theta\,\left\{\frac{1}{4}{\mcN}_{AB}(\Sigma)\mcW^A\mcW^B+3\mcN(V_y)^{\frac{1}{3}}\left[h^+e^{-g_MV^M}h+h^{c+}e^{g_MV^M}h^c\right]^{\frac{2}{3}}\right\}\\
	       &\qquad-2\int d^2\theta \,h^c(\partial_y-g_M\Sigma^M)h+\textup{h.c.},
\end{split}\ee
where the combination $g_M{\mathbb V}^M$ gauges an U(1)$_R'$ subgroup of the SU(2)$_R$ which (unlike the U(1)$_R$ of the previous paragraphs) survives the orbifold projection. The constants $g_M$ are Fayet-Iliopoulos couplings. Notice that there is another type of FI couplings, namely odd FI couplings , which don't survive upon reduction to 4D (see \cite{us04b} and references therein).  

The \emph{na\" ive} effective 4D couplings, obtained from \eqref{eq:FIterms1}, are determined by the K\"ahler potential $\mcK=-\ln{\tilde\mcN}(\Sigma+\Sigma^+)$, and the D-term potential
\be
                    V_D=\frac{1}{4}{\hat\mcN}_{NM}(\Sigma+\Sigma^+)D^ND^M-\ka_4^{-2}{\bar g}_MD^M.
\ee
There is no superpotential and thus also no contribution of F-terms to the potential. After integrating out the $D^M$ we finally get 
\be\label{eq:vidi}
                 V_D=-M_P^4{\bar g}_N{\hat\mcN}^{NM}{\bar g}_M.
\ee
Even though it resembles the F-term potential in eq.\eqref{eq:FtermPot1}, this potential displays a crucial difference: There is no such term like the second one in \eqref{eq:FtermPot1}, which in this case would read
\be
                      V\supset-\frac{M_P^4}{2\tilde\mcN}({\bar g}_M(\Sigma^M+\Sigma^{M+}))^2.
\ee 
The reason is that in our reduction to a 4D effective theory we dropped the odd fields $\Sigma^M$, otherwise there would be such a term as the one above. As this term gives \emph{negative} squared masses to the $\Sigma^M$ it is clear that even though these are \emph{odd} superfields, they cannot be consistently set to zero unless the FI terms ${\bar g}_M$ are small enough so that the Kaluza-Klein mass compensates their effect. However, for a large enough radius of the 5th direction the $\Sigma^M$ will again become tachyonic. This is something one should worry about.

Let's come back to the D-term potential $V_D$ of \eqref{eq:vidi} and discuss its possible use for moduli stabilization. We consider as an example again the norm function $\ka_5\mcN=(\sigma_0)^3-\sigma_0(\sigma_1)^2$, but now $\sigma_0\sim+$ while $\sigma_1\sim -$. We get the following potential
\be
                    V_{(D)}=\frac{1}{2}\frac{M_P^4{\bar g}_1^2}{\sigma_0},
\ee 
which is essentialy the same as $V_{(B)}$ stripped off of the tachyonic field $\psi$, which now is an odd field that we assume to be stabilized by its KK mass. Unlike the F-term potentials, $V_{(D)}$ induces a runaway behaviour for the radion $\sigma_0$ towards decompactification. In their work on dS flux compactifications \cite{KKLT}, KKLT proposed to use such a D-term potential to uplift the supersymmetric AdS vacua induced by non-perturbative superpotentials. It is by now clear that this cannot be done in a way consistent with the gauged $R$-symmetry \cite{Choi05,deAlwis05,Villadoro05,Freedman:2005up}. In fact, under the above U(1)$_R'$, the compensator transforms \cite{Ferrara:1983dh} as $h\to e^{g_M\Lambda^M}h$, and accordingly any superpotential $W$ (defined in eq.\eqref{eq:def_superpotent}) will have to transform as $W\to e^{-2g_M\Lambda^M}W$. This puts a non-trivial constraint on the form of the allowed superpotentials. In particular, superpotentials of the type $W=a+be^{-c\Sigma}$, with constant $a,b,c$ are not compatible with FI terms (of the above type). 

So, it seems that it is fair to say that FI terms are problematic for moduli stabilization. This is true, in particular, for the \emph{first} of the two stabilization mechanisms discussed in ref.\cite{dudas05}, which uses an FI term to give mass to a vector multiplet. We will have more to comment on this in sec.\ref{subsec:kaehlerstuff}.

\subsection{One-loop effects and moduli stabilization}\label{subsec:kaehlerstuff}

As it turns, one-loop corrections to the K\" ahler potential, due to bulk multiplets, can lead to moduli stabilization in ${\mathbb M}_4$ vacua \cite{Ponton:2001hq,Luty:2002hj}. A more detailed discussion of this kind of stabilization mechanism was then presented in refs.\cite{vonGersdorff:2003rq,dudas05} (see \cite{vonGersdorff:2005ce} for a discussion of two-loop effects), for the single modulus (i.e. the radion) case. Here we will present the supergravity embedding of these models, generalizing them to the many moduli case, and consider \emph{new} one-loop corrections to $\mcK(\Sigma,\Sigma^+)$. These new corrections will be crucial to lift the flatness of the potential in the axionic directions, as they break the continuous \emph{shift symmetry} (common for no-scale type theories) present at tree-level.

There are essentially two kinds of loop corrections which one might call moduli-independent or moduli-dependent, respectively. The first depends only on the combination of the moduli which measures the size of the extra-dimension, the radion. The second one depends also on other combinations of the moduli and arise when some multiplets are charged under some bulk abelian gauge symmetries (that is, some isometries of the scalar manifold are gauged by vector multiplets). These two kinds of corrections correspond to the contributions of massless and massive bulk multiplets of rigid supersymmetry.

\begin{center}
{\bf Moduli-independent one-loop contributions}
\end{center} 
We consider first the contributions of the gravitational sector, abelian vector multiplets in the unbroken phase, and uncharged hypermultiplets. These lead to a 1-loop correction to the low-energy D-term Lagrangian of the form,
\be
                    \Delta \mcL_V=-\int d^4\theta\frac{\alpha}{R^3{\mathbb W}_y^2}+(\textup{higher powers in a superderivative expansion}),
\ee
where $\alpha=(-2-N_V+N_H)\frac{\zeta(3)}{2(2\pi)^5}$ depends on the number $N_V$ of "massless" vector multiplets and on the number $N_H$ of "massless" hypermultiplets. Integrating out ${\mathbb W}_y$, eq.\eqref{eq:wy} is modified and one finds the following one-loop K\" ahler potential\footnote{See also \cite{Rattazzi:2003rj,Falkowski:2005fm} and references therein.}
\be\label{eq:kaehler_with_corrections}
        \mcK(\Sigma,\Sigma^+)= -\ln\left({\tilde\mcN}(\Sigma+\Sigma^+)+\Delta\right), 
\ee
where
\be\label{eq:kaehler_corrections_massless}
                   \Delta=\Delta_{\alpha}\equiv\ka_5^2\frac{\alpha}{R^3}.       
\ee          
For a detailed derivation (in superfield language) of this and other type of 1-loop corrections to the K\"ahler potential see \cite{usnew}.

\begin{center}
{\bf Moduli-dependent one-loop contributions} 
\end{center}
There are several ways of gauging isometries of the scalar manifold and therefore there are also different possible 1-loop corrections to $\mcK(\Sigma,\Sigma^+)$: 

Let us start with a physical hypermultiplet ${\mathbb H}=(H,H^c)\sim(+,-)$ charged under the vector multiplet ${\mathbb V}=(V,\Sigma)\sim(-,+)$. The relevant couplings are
\be\label{eq:hyper_phys_charged_1}
         \int d^2\theta \,(2H^c\partial_y H-g_{\beta}\Sigma (H^2-H^{c2}))+\textup{h.c.},     
\ee
and the corresponding one-loop contribution to $\mcK(\Sigma,\Sigma^+)$ in eq.\eqref{eq:kaehler_with_corrections}, due to hypermultiplets running in the loop is
\be\label{eq:Delta_hyper_charged_1}
                 \Delta_{\beta}=\kappa_5^2\frac{\beta}{R^3}(e^{-2\pi Rg_{\beta}\Sigma}+e^{-2\pi Rg_{\beta}\Sigma^{+}})\left(1+\pi Rg_{\beta}(\Sigma+\Sigma^{+})\right)+\textup{higher }k,
\ee
with $\beta=N_{H_{\beta}}/{4(2\pi)^5}$, where $N_{H_{\beta}}$ is the number of charged hypermultiplets. In the most general case $g_{\beta}\Sigma$ can be replaced by a linear combination of moduli $g_I\Sigma^I$. A feature which we did not display is that \eqref{eq:Delta_hyper_charged_1} is only valid for $\textup{Re}(\Sigma)>0$, for $\textup{Re}(\Sigma)<0$ we must replace $\Sigma\to-\Sigma$. There are several ways of deriving eq.\eqref{eq:Delta_hyper_charged_1}, the most elegant one using the relation of the K\"ahler potential of $\mcN=2$ supergravity to the prepotential, $\mcK=-\ln\left(-2(\mcF+\mcF^+)+(\mcF_I+\mcF_{\bar I}^+)(\Sigma^I+\Sigma^{{\bar I}+})\right)$. This and also a supergraph derivation will be presented in \cite{usnew}.

At this point one should make a remark regarding the regime of validity of the above 1-loop contribution to the effective K\" ahler potential. In fact it relies on $\Sigma$ being large enough as to make the lowest lying hyper KK mode massive enough compared to $\Sigma$ it-self. If this is the case, it implies in turn that there is no effective superpotential arising from \eqref{eq:hyper_phys_charged_1}. This is indeed the case in specific models that we will consider in \cite{usnew}.\\ 

Another type of coupling between a hypermultiplet and a vector multiplet with the same orbifold parities as in the previous example, is
\be
          \int d^2\theta \,2H^c(\partial_y-g_{\gamma}\epsilon(y)\Sigma) H+\textup{h.c.},            
\ee   
where $\epsilon(y)=\partial_y|y|$ is the periodic step-function. Integrating out the hypermultiplets we get the following contribution to $\Delta$
\be\label{eq:Delta_hyper_charged_2}
         \Delta_{\gamma}=\kappa_5^2\frac{\gamma}{R^3}e^{-\pi Rg_{\gamma}(\Sigma+\Sigma^+)}\left(1+\pi Rg_{\gamma}(\Sigma+\Sigma^{+})\right)+\textup{higher }k,    
\ee
where $\gamma=N_{H_{\gamma}}/{2(2\pi)^5}$ and $N_{H_{\gamma}}$ is the number of hypermultiplets. This expression can e.g. be obtained from formulae given in \cite{vonGersdorff:2003rq} or by a supergraph calculation \cite{usnew}. A relevant difference to the case discussed previously lies in the fact that the lowest KK mode is now massless before supersymmetry is broken \cite{vonGersdorff:2003qf}. It might be present in the low-energy field content. But the \emph{crucial} difference to eq.\eqref{eq:Delta_hyper_charged_1} is that it doesn't depend on the axionic modulus $\textup{Im}(\Sigma)$.

Even though $\Delta_{\beta}$ and $\Delta_{\gamma}$ have a similar form, in the presence of a constant superpotential $W_0$ they lead to rather different potentials. We find that by a suitable arrangement of the parameters $\alpha$, $\beta$ and $\gamma$ as well as of the couplings $g_{\beta}$ and $g_{\gamma}$ it is possible to obtain potentials with ${\mathbb M}_4$ or dS \emph{minima} \cite{talk}. Notably, at these minima also the axionic part of the moduli are stabilized. The detailed study of these potentials will be presented elsewhere \cite{usnew}.

Finally, we would like to comment on the two models of radion stabilization presented in \cite{dudas05}. An important ingredient of these models is a 1-loop correction due to "massive" bulk vector multiplets. In addition to the fact that it doesn't stabilize the axionic moduli, in our view the \emph{first} of these models has the problem that the vector multiplet gets its mass from spontaneous symmetry breaking induced by a Fayet-Iliopoulos term. As we already pointed out, it is difficult to conciliate an FI term with a \emph{constant} superpotential and therefore, as it stands, we think the first model of \cite{dudas05} is not viable for moduli stabilization. On the other hand, the second model, involving the Hosotani breaking of a bulk SU(2) gauge symmetry, does not share these problems. It would be interesting to embbed this model in 5D orbifold supergravity, however this goes beyond the scope of this paper.

\subsection{Non-perturbative effects from gaugino condensation}

As we have already mentioned, the flatness of the potential will be lifted 
either if the K\"ahler potential receives non cubic (with respect to 
$\Si^I+\Si^{I+}$) corrections, or the superpotential has a moduli 
dependent 
part. In this subsection we will show how this can be 
achieved by gaugino condensation. 
In this case the effective superpotential will have a moduli dependent 
(non perturbative) part. Also the K\"ahler potential gets a non 
perturbative correction. The whole effective action can still be
written in a superconformal form
\beq
-3\int d^4\theta\, e^{-{\cal K}/3}\phi^+\phi+\int d^2\theta\, \phi^3\,W + {\rm h.c.}
\la{efact}
\eeq
Discussing gaugino condensation, we follow an effective 4D description
dealing with the zero modes of the relevant states. 
Thus, the arguments applied for a pure 4D gaugino condensation 
scenario will be appropriate.
For this case the super and K\"ahler potentials include perturbative (p) and 
non-perturbative (np) parts \cite{Burgess:1995aa}:
\beq
W=W_p+W_{np}~,~~~e^{-{\cal K}/3}=e^{-{\cal K}_p/3}-k\,e^{-{\cal K}_{np}/3}~,
\la{efpots}
\eeq
where $k$ is some constant. The action with (\ref{efpots}) is valid below
a scale $\La $ corresponding to the energy scale at which the gauge sector becomes strongly coupled.

{}For demonstrative purposes we will 
consider an example with one non-Abelian $SU(N)$ YM theory which is
responsible for the gaugino condensation. 
Taking the norm function
\beq
{\cal N}(M)=M_0^3-M_0{\rm Tr}(M_g)^2~, ~~~
{\rm with}~~~M_g=\fr{1}{2}\lam^aM^a~,~~~a=1,\cdots , N^2-1~,
\la{NM}
\eeq
the coupling of the moduli superfield $\Si^0$ with the gauge field strength
will be
\beq
-\fr{1}{4}\int d^2\theta\, {\cal N}_{IJ}(\Si ){\cal W}^I{\cal W}^J+{\rm h.c.}
\to \fr{1}{4}\int d^2\theta\, \Si^0{\cal W}^a{\cal W}^a+{\rm h.c.}
\la{SiWW}
\eeq
The effective 4D superconformal theory with (\ref{efact}), (\ref{SiWW})
possesses Weyl and chiral $U(1)$ symmetries at classical level, a superposition of which is anomalous on the
quantum level. Namely, a mixed gauge-chiral anomalous term is 
generated and the counter term \cite{Derendinger:1991hq,Kaplunovsky:1994fg}
\beq
-2c\int d^2\theta\, \fr{1}{4}\ln \phi {\cal W}^a{\cal W}^a+{\rm h.c}~
\la{conterm}
\eeq
is needed to take care of the anomaly cancellation. The coefficient $c$
in (\ref{conterm}) is related to the gauge group $b$-factor and is positive 
for asymptotically free theories.

Therefore, the coupling of the composite chiral superfield 
$U=\lan {\cal W}^a{\cal W}^a\ran $ with moduli and compensating superfields
is given by
\beq
\fr{1}{4}\int d^2\theta\, (\Si^0-2c\ln \phi )\,U +{\rm h.c.}
\la{consup}
\eeq 
Eq. (\ref{consup}) is the starting
point for computing the non perturbative effective action $\Ga (\Si^0, U)$.
After obtaining $\Ga (\Si^0, U)$, one can minimize it with 
respect to $U$ (determining the condensate $U_0$) and then plugging
back the value of $U=U_0$ in $\Ga (\Si^0, U)$  derive the effective
action for the $\Si^0$-modulus. In the case of a single condensate, the non-perturbative K\"ahler and superpotentials are given by \cite{Burgess:1995aa}
\beq
{\cal K}_{np}=\fr{3}{2c}(\Si^0+\Si^{0+ })~,~~~~
W_{np}=\tl{W}\exp\left(-\fr{3\Si^0}{2c}\right)~.
\la{nonp}
\eeq
It is interesting to note the similarity of the non-perturbative superpotential $W_{np}$ with the tree-level ones of section \ref{sub:FIterms} in the small modulus limit $\Sigma^0\to 0$. As is the case for the tree-level superpotentials, with the addition of a constant brane superpotential it is possible to obtain AdS vacua and therefore, also in this case, an additional
contribution must be generated \cite{luty99} in order to set the vacuum energy to zero.

The tree level perturbative K\"ahler potential has the form 
${\cal K}_p=-3\ln (\Si^0+\Si^{0+ })$. Therefore the total 
K\"ahler potential
will be
\beq
{\cal K}=-3\ln \l \Si^0+\Si^{0+ }-
ke^{-\fr{\Si^0+\Si^{0+ }}{2c}}\r ~.
\la{totK}
\eeq
Note that the non-perturbative correction to the K\"ahler potential in the 
problem of moduli stabilization is usually not taken into account in the literature. In fact, there is no physical reason to exclude the $k$-term from considerations. Since the coefficient $k$ is related to a mass scale, one should expect that it
will have an impact on the stabilized value of the modulus $\Si^0$. Indeed, we find that the inclusion of the non-perturbative 
part of the K\"ahler potential can introduce significant effects in the stabilizing potential. A detailed study of the effects of gaugino condensation and their interplay with the perturbative one-loop contributions in the context of moduli stabilization will be presented elsewhere \cite{usnew}.


\locsection{Reduction to 4D in warped geometries}\label{sec:warp}

This section deals with the implementation of the approach developed above now for warped geometries. We want to take into account a possible gauging of the U(1)$_R$ symmetry with an \emph{odd} coupling\footnote{See \cite{Bagger:2002rw} where the relation between the 'odd coupling' formulation and the'even coupling' formulation of \cite{Altendorfer:2000rr} is discussed. The ABN version \cite{Altendorfer:2000rr} with an even gauge coupling leads
to 'detuned' brane tensions. This was discussed in ref. \cite{Kugo:2002js}
and in ref. \cite{Bagger:2002rw} and subsequent papers. The detuned version
in \cite{Bagger:2002rw} leads to ${\rm AdS}_5$ in the bulk and to spontaneously broken SUSY on the branes.}. As it was pointed out in \cite{Gherghetta:2000qt,Falkowski:2000er,Bergshoeff:2000zn} and in superfield description developed in \cite{us04a,us04b}, this is needed to obtain the SUSY Randall-Sundrum model \cite{Randall:1999ee} and its generalizations. Again, for simplicity, at first we do not consider physical hypermultiplets. The hyper Lagrangian in this case is (see \cite{us04a})\footnote{We refer the reader to appendix D and section 4 of \cite{us04a} for a detailed explanation of the Lagrangians appearing in this section. This includes a derivation of the bosonic part of the component action, as well as of the BPS equations of motion which fix the warped geometry dynamically.}
\be\label{eq:Dlagran_warp1}
              \mcL=-3\int d^4\theta \,\mcN(V_y)^{\frac{1}{3}}\left[h^+e^{-\frac{3\epsilon}{2}k_IV^I}h+h^{c+}e^{\frac{3\epsilon}{2} k_I V^I}h^c\right]^{\frac{2}{3}}-2\left(\int d^2\theta \,h^c(\partial_y - \tfrac{3}{2}\epsilon(y)k_I\Sigma^I)h+\textup{h.c.}\right).\vspace{6pt}
\ee

We will first show how to obtain the low-energy Lagrangian for the single modulus case, i.e. the only modulus is the radion superfield $\Sigma^0$. This is the supersymmetric RS-model. For purposes that will become clear below, we introduce the chiral superfield $\phi$ as
\be\label{eq:wave_comp_even}
                h=\ka_5^{-1}\exp{\left(\frac{3k_0}{2}\int_0^y dy\epsilon(y)\Sigma^0\right)}\,\phi^{\frac{3}{2}},
\ee 
and similarly
\be\label{eq:wave_comp_odd}
                h^c=\ka_5^{-1}\exp{\left(-\frac{3k_0}{2}\int_0^y dy\epsilon(y)\Sigma^0\right)}\,\phi^{\frac{3}{2}}\phi^c
\ee 
to get\footnote{Here and in the following we assume that the odd vector superfield $V$ vanishes at the boundaries, i.e. $V(0)=V(\pi R)=0$. This would not be the case in the presence of brane localized kinetic terms. In section \ref{eq:boundary} we explain how to handle that situation.}
\be\label{eq:Dlagran_warp2}
             \mcL=-3\ka_5^{-1}\int d^4\theta\, V_y^0\,e^{k_0\int_0^y dy\epsilon \,V_y^0}\phi^+\phi\left[1+\phi^{c+}\phi^c\,e^{-3k_0\int_0^y dy\epsilon\, V_y^0}\right]^{\frac{2}{3}} -\left(\int d^2\theta \,\ka_5^{-2}\phi^c\partial_y\phi^3 +\textup{h.c.}\right)+\cdots
\ee
The \emph{crucial} point now is that in a supersymmetric compactification the chiral compensator $\phi$ cannot depend on $y$, the fifth coordinate. This fact follows from the BPS condition $F_{\phi^c}^*\propto\partial_y\phi^3=0$ which is obtained \cite{us04b} from \eqref{eq:Dlagran_warp2}. We see that all the $y$-dependence of the compensator $h$ is entailed in the exponential $e^{\int^y dy\epsilon(y)\Sigma^0}$, implying that in the $e^5_y=1$ \emph{gauge} 
\be
                  h=\ka_5^{-1}\exp\left(\frac{3}{4}\ka_5^{-1}\,k_0\,|y|\right)+\cdots
\ee
Thus, according to \eqref{eq:fixing_1} (for $|\phi_H|^2=|\phi_H^c|^2=0$) there is a non-trivial warp-factor of RS-type, given as $2\sigma(y)=\ka_5^{-1}\,k_0\,|y|$.

Now, we see that it is possible to perform the integration of \eqref{eq:Dlagran_warp2} over $y$ exactly, since\vs{4pt}
\be\label{eq:81}
            V_y^0\,\exp\left(k_0\int_0^y dy\epsilon\,V_y^0\right)=         \frac{\epsilon(y)}{k_0}\partial_y \exp\left(k_0\int_0^y dy\epsilon\,V_y^0\right),\vs{4pt}   
\ee
and $\partial_y\phi =\phi^c=0$. Using this we are able to obtain the effective 4D Lagrangian in a straightforward way. We get\footnote{For notational simplicity we will sometimes write $\oint=\int_{-\pi R}^{\pi R}$.}
\be
            \mcL^{(4D)}=\int_{-\pi R}^{\pi R}\,dy\mcL=-6\ka_5^{-1}\int d^4\theta\frac{1}{k_0}\left(e^{\frac{k_0}{2}\oint dy(\Sigma^0+\Sigma^{0+})}-1\right)\phi^+\phi,
\ee
which in terms of the radion chiral superfield $T\equiv\frac{1}{2\pi}\oint dy\Sigma^0$ (and with $k=k_0$) becomes
\be
          \mcL^{(4D)}=-6\ka_5^{-1}\int d^4\theta\frac{1}{k}\left(e^{k\pi(T+T^+)}-1\right)\phi^+\phi,  
\ee
the very same expression obtained in refs.\cite{Luty:2000ec,Bagger00}. The corresponding K\" ahler potential is
\be\label{eq:RS_kaehler}
             \mcK_{RS}(T,T^+)=-3\ln2\frac{e^{k\pi(T+T^+)}-1}{k}.
\ee
Note that to obtain this expression no detailed knowledge on the $x^{\mu}$ and $y$ dependencies of the radion $\Sigma^0$ or of $V^0$ was needed. In fact, it is known \cite{Luty:2000ec,Bagger00} that for $\partial_{\mu}\textup{Im}(\Sigma^0)\neq 0$ a \emph{tadpole} leads to a non-zero VEV of the superfield $\partial_y V^0$. This has the crucial effect of ensuring the supersymmetry of the low-energy action. But, in the superfield approach we use here we do not need to know the exact value of $\partial_y V^0$, since  
\be\label{eq:4.9}
                          \oint dy\, V_y^0=\oint dy\,(\Sigma^0+\Sigma^{0+})=2\pi(T+T^+).
\ee

\subsection{General warped geometries without hypermultiplets}

Clearly, the derivation of the K\"ahler potential for the supersymmetric RS model was highly simplified by the assuption of only one existing modulus. What happens in the more general case with multiple vector moduli? Looking back to eq.\eqref{eq:Dlagran_warp1} we introduce now the 4D compensator chiral superfield as
\be
                h=\ka_5^{-1}\exp{\left(\frac{3}{2}\int_0^y dy\epsilon(y)k_I\Sigma^I\right)}\,\phi^{\frac{3}{2}}.
\ee 
It is then clear that the effective 4D K\"ahler potential should be obtained by evaluating the following expression
\be\label{eq:master_kaehler}
                \mcK=-3\ln\left\{\oint dy\,\mcN^{\frac{1}{3}}(\Sigma+\Sigma^+)\,\exp\left(\int^y_0dy\,\epsilon(y) k_I(\Sigma^I+\Sigma^{I+})\right)\right\}.
\ee
Unfortunatly it is in general not straightforward to calculate this explicitely, the only simple case being the one we already discussed. There are two reasons for this difficulty: the first complication is that $\Sigma^I$ depends both on $x^{\mu}$ and $y$, the second one is that in most cases we are not even able to explicitely know what this dependence is. Only in the single modulus case it is possible to obtain a closed expression for $\mcK$ without relying on any knowledge on the $(x^{\mu},y)$-dependence of $\Sigma^0$. 

On the other hand, there is a piece of information which we can use to evaluate \eqref{eq:master_kaehler} within some level of approximation as we will explain now. To be precise, we can explore the fact that \eqref{eq:master_kaehler} is invariant under a shift of the superfield $\Sigma^I$ by an imaginary constant, $\Sigma^I\to\Sigma^I+ic^I$. For the moduli $T^I$ defined as
\be
             T^I\equiv\frac{1}{2\pi}\oint dy\Sigma^I,
\ee
this shift symmetry becomes $T^I\to T^I + iRc^I$. It follows that the K\"ahler potential $\mcK$ can only be a function of $T^I+T^{I+}$, and eventually of superfield derivatives of $T^I\pm T^{I+}$. But the latter contribute only to higher order space-time derivative terms in the Lagrangian and can be droped in a truncation at two-derivative order. We will in the following work at the level of such a truncation. Then, by supersymmetry, the dependence of $\mcK$ on $T^I+T^{I+}$ can be obtained just by considering the real part $t^I$ of $T^I$'s lowest component, and using the replacement $2t^I\to T^I+T^{I+}$ at the end of the calculation.  


We start with the following 5D metric
\be\label{eq:ansatz}
              ds_5^2=e^{2\sigma}dx_4^2+e^{-4\sigma}dy^2.                      
\ee
As we just said, to obtain an explicit expression for the K\"ahler potential \eqref{eq:master_kaehler}, we only need to know the explicit $(x^{\mu},y)$-dependence of $\Sigma^I$ ($I=0,1$) if we want to go beyond the two-derivative truncation. We should note, however, that the above metric is expected (cf. \cite{brax04}) to be a good \emph{Ansatz} also for "small" fluctuations around the $x$-independent solution of the BPS equations of motion, i.e. when we replace $\sigma(y)\to\sigma(x,y)$. 

Now, the BPS equations can be obtained by requiring vanishing D-terms and F-terms \cite{us04b} and read 
\be\label{eq:D_flat}
                 \partial_y\left(e^{2\sigma}\mcN_I(M)\right) = 3\epsilon(y)M_5^3\,k_I,
\ee
(where we already used the gauge $e^5_y=e^{-2\sigma}$) and
\be\label{eq:for_warp}
                  (\partial_y-e^5_y\epsilon(y)k_IM^I)e^{2\sigma}=0. 
\ee
With \eqref{eq:ansatz} and also using eq.\eqref{eq:for_warp} one readily finds a simple expression for the K\"ahler potential\footnote{Note that despite the fact that $\mcK$ is obtained by evaluating $\sigma(y;T+T^+)$ at the $y=0$ brane, there is no asymmetry between the $y=0$ brane and the $y=\pi R$ brane. In fact, $\sigma(0;T+T^+)$ and $\sigma(\pi R;T+T^+)$ differ just by a K\"ahler gauge transformation.} 
\be\begin{split}\label{eq:kaehler_warp}
              \mcK(T,T^+) & = -3\ln\left\{\oint dy\,e^5_y\,\exp\left(2\sigma(y)-2\sigma(0)\right)\right\}=\\
	                  & =6\,\sigma(\,0\,;T^I+T^{I+})-3\ln2\pi R,
\end{split}\ee
where $\sigma(\,0\,;T^I+T^{I+})$ is the warp-factor evaluated at $y=0$, as obtained by solving eqs.\eqref{eq:D_flat}, written in terms of $T^I+T^{I+}=2t^I$. 

Let us explain this with the following particular example: we consider the setup defined by the norm function $\mcN(M)=\ka_5(M^0)^2 M^1$ and the charges $k_0,k_1\neq 0$. Then the eqs.\eqref{eq:D_flat} are solved with
\be\label{eq:warpfac}
                       e^{2\sigma}=\ka_5^{-1}\left(\frac{3k_0}{2}|y|+\frac{b_0}{2}\right)^{\frac{2}{3}}\left(3k_1|y|+b_1\right)^{\frac{1}{3}},
\ee
and
\be
                e^5_yM^0=\frac{2}{3k_0|y|+b_0},\qquad e^5_yM^1=\frac{1}{3k_1|y|+b_1}.
\ee
(The $x$-dependence would enter these expressions, and in particular $\sigma(x,y)$, through the integration constants $b_I$ which would be promoted to $x$-dependent quantities $b_I(x)$ (see e.g.\cite{brax04}).) It is then not difficult to see that 
\be
               b_0=\frac{3k_0\pi R}{e^{3k_0\pi t^0}-1},\qquad  b_1=\frac{3k_1\pi R}{e^{6k_1\pi t^1}-1}.
\ee
Finally, plugging this back in \eqref{eq:warpfac} and then in eq.\eqref{eq:kaehler_warp} we obtain the K\"ahler potential as 
\be\label{eq:kaehler_special}
           \mcK(T,T^+)=-2\ln4\frac{e^{\frac{3}{2}k_0\pi (T^0+T^{0+})}-1}{3k_0}-\ln2\frac{e^{3k_1\pi (T^1+T^{1+})}-1}{3k_1}.
\ee
Clearly, $\mcK$ has the correct behaviour in the unwarped limit $k_0,k_1\to 0$, that is $\mcK\to-\ln{\tilde\mcN}(T+T^+)$.

Here, a few remarks are in order. Let us first stress again that the relation $\mcK(T,T^+)=6\,\sigma(\,0\,;T^I+T^{I+})$ was obtained under the assumption that truncating the Lagrangian at second order in the derivatives is a good approximation. In addition, to obtain the K\"ahler potential of eq.\eqref{eq:kaehler_special} we had to rely on our ability to write the warp-factor $\sigma(y)$ in terms of the moduli $t^I=(T^I+T^{I+})/2$. In more general cases this will not be possible, at least not in an exact way. Then one will have to work within additional approximations to obtain explicit expressions for $\mcK(T,T^+)$. 

It is interesting to note that also in ref.\cite{brax04} a general expression for the effective K\"ahler potential for gauged SUGRA was presented, which was then evaluated explicitely for the model we just considered above. It would be nice to understand if and how their expression is related to ours (eq.\eqref{eq:kaehler_special}), but in fact we were not able to find such a relation (beyond the unwarped limit)\footnote{After concluding this work, we noticed a subtle error in eq.(3.14) of ref.\cite{brax04}. This error propagates throughout the paper in such a way that in ref.\cite{brax04} what is called a K\"ahler potential is in fact the so-called \emph{sympletic potential}, which is related to the K\"ahler potential by a Legendre transform.
For a recent description of both the complex and the sympletic
formulations of K\"ahler geometry see \cite{Martelli:2005tp}.}
. 

An additional example of an exactly solvable model is the so-called STU-model, which is defined by $\mcN(M)=\ka_5 M^0 M^1 M^2$ (see \cite{Bergshoeff:2000zn}). In this case we find that the K\"ahler potential reads 
\be\label{eq:kaehler_STU}
           \mcK_{STU}(T,T^+)=-\sum_{I=0}^2\ln2\frac{e^{3k_I\pi (T^I+T^{I+})}-1}{3k_I}.
\ee
Needless to say, also the simple RS-model of eq.\eqref{eq:RS_kaehler} can be recovered in the same way.

\subsection{Hypermultiplets and supersymmetric Minkowski vacua}\label{sec:susystab}

Let us consider now the addition of one \emph{physical} hypermultiplet $(H,H^c)\sim(+,-)$ also charged under the U(1)$_R$ with odd coupling. To take this into account, the 5D Lagrangians undergo the following modifications 
\be
              \mcL_D=-3\int d^4\theta \,\mcN(V_y)^{\frac{1}{3}}\left[h^+e^{-\frac{3\epsilon}{2}k_IV^I}h+h^{c+}e^{\frac{3\epsilon}{2} k_I V^I}h^c-H^+e^{-\frac{3\epsilon}{2}q_IV^I}H+H^{c+}e^{\frac{3\epsilon}{2} q_I V^I}H^c\right]^{\frac{2}{3}},
\ee
\vspace{2pt}
\be
              \mcL_F=-2\int d^2\theta \,h^c(\partial_y - \tfrac{3}{2}\epsilon(y)k_I\Sigma^I)h+2\int d^2\theta \,H^c(\partial_y - \tfrac{3}{2}\epsilon(y)q_I\Sigma^I)H+\textup{h.c.}\vspace{6pt}
\ee
For the sake of simplicity we just consider the case with $\mcN=\ka_5 (M^0)^3$. More general scalar manifolds can in priciple be handled using \eqref{eq:kaehler_warp}. Due to the global symmetry $\Phi\to\Phi\,e^{i\alpha}$, the K\"ahler potential can be determined at two-derivative order just by considering a constant value $\varphi$ of the scalar component of $\Phi$, and using the replacement $|\varphi|^2\to \Phi^+\Phi$ at the end of the calculation. 

To obtain the low-energy action we have, again, to factor out the y-dependence of the superfields. In addition to eqs.\eqref{eq:wave_comp_even} and \eqref{eq:wave_comp_odd}, we have for the physical hypers 
\be
                H=\ka_5^{-1}\Phi_0\exp{\left(\frac{3q_0}{2}\int_0^y dy\epsilon(y)\Sigma^0\right)}\,\phi^{\frac{3}{2}},
\ee 
and
\be
                H^c=\ka_5^{-1}\Phi_0^c\exp{\left(-\frac{3q_0}{2}\int_0^y dy\epsilon(y)\Sigma^0\right)}\,\phi^{\frac{3}{2}}.
\ee 
In terms of the new variables, the F-term Lagrangian reads
\be
              \mcL_F=-\ka_5^{-2}\int d^2\theta \,\left(\phi^c\partial_y\phi^3+\phi^3(\Phi_0\partial_y\Phi_0^c-\Phi_0^c\partial_y\Phi_0)\right)+\textup{h.c.}
\ee
Again, supersymmetry requires that $\partial_y\phi=\partial_y\Phi_0=0$. This follows from the BPS conditions $F_{\phi^c}=F_{\Phi_0^c}=0$. Moreover, unless we turn on a brane localized superpotential, $F_{\Phi}\propto \partial_y\Phi_0^c=0$ implies the odd superfield $\Phi_0^c$ to vanish at low energies. The 4D Lagrangian is thus obtained by integrating\footnote{Note that, according to \eqref{eq:fixing_1}, for a non-vanishing hypermultiplet ($|\Phi_0|^2\neq 0$) the warp-factor is $2\sigma(y)=\ka_5^{-1}\,k_0|y|+\ln\left\{1-|\Phi_0|^2\exp\left(\frac{3}{2}(q_0-k_0)|y|\right)\right\}$, in the $e^5_y=1$ gauge. This visibly differs from the RS case.}
\be
               \mcL=-3\ka_5^{-1}\int d^4\theta \,V_y^0\,e^{k_0\int_0^y dy\epsilon \,V_y^0} \phi^+\phi\left[1-\Phi_0^+\Phi_0\,e^{\frac{3}{2}(q_0-k_0)\int_0^y dy\epsilon\, V_y^0}\right]^{\frac{2}{3}},
\ee
over the $S^1/{\mathbb Z}_2$ orbifold. Proceeding like in \eqref{eq:wave_comp_even} ff it is straightforward to obtain the corresponding K\"ahler potential
\be\label{eq:rad_hyp}
                      \mcK(T+T^+,|\Phi_0|^2)=-3\ln \frac{2}{k_0}\int_1^{e^{\pi k_0 (T+T^+)}} \hspace{-12pt}d\rho\,\left[1-|\Phi_0|^2\,\rho^{\frac{3}{2}\left(q_0/k_0-1\right)}\right]^{\frac{2}{3}},
\ee 
which for $|\Phi_0|^2=0$ reduces to $\mcK_{RS}$ (cf. eq.\eqref{eq:RS_kaehler}), as it should. It is interesting to consider the special case $3q_0=5k_0$, where we can evaluate $\mcK$ exactly
\be\begin{split}\label{eq:rad_hyp_simp}
                  \mcK_{3q_0=5k_0} & =-3\ln \frac{6}{5k_0|\Phi_0|^2}\left\{\left(1-|\Phi_0|^2\,e^{\pi k_0 (T+T^+)}\right)^{\frac{5}{3}}-\left(1-|\Phi_0|^2\right)^{\frac{5}{3}}\right\}\\[12pt]
		  & = -3\ln\left\{\left(1-|\Phi_{\pi}|^2\right)^{\frac{5}{3}}-\left(1-|\Phi_0|^2\right)^{\frac{5}{3}}\right\}+\cdots
\end{split}\ee
Here we introduced $\Phi_{\pi}\equiv\Phi_0\,e^{\pi k_0 T}$ and the ellipsis stand for terms which can be gauged away by a K\"ahler transformation. Note that $\Phi_0$ and $\Phi_{\pi}$ are the values that the 5D superfield $\Phi(y)=H/h$ takes at the two branes. 

Looking back at eq.\eqref{eq:rad_hyp_simp}, one observes an intricate mixing of the moduli in the K\"ahler potential, that is typical for warped compactifications. The same kind of structure also appears in the reduction of 5D heterotic M-theory, that we will perform in the following section. This has some interesting consequences, in particular it turns out that certain "dualities" that would exist in the unwarped limit $k_0=q_0=0$, namely $T\to T^{-1}$ and $\Phi\to\Phi^{-1}$, are \emph{absent} (or have to be modified) in the warped case with $\Phi_i\neq 0$. On the other hand, in the limit that $\Phi_0\to 1$, which corresponds to a Randall-Sundrum II like model with $e^{2\sigma(0)}=0$, we get $\mcK\to-5\ln(1-|\Phi_{\pi}|^2)$, which displays the "duality" $\Phi_{\pi}\to\Phi_{\pi}^{-1}$. Similar properties appear in the heterotic M-theory case, see eq.\eqref{eq:limit2}.  

Another pertinent observation is that also for arbitrary values of $q_0$ and $k_0$ the K\"ahler potential of eq.\eqref{eq:rad_hyp}, written as a function of $\Phi_0$ and $\Phi_{\pi}=\Phi_0\,e^{\frac{3}{2}(q_0-k_0)\pi T}$, has a rather suggestive form:
\be
                      \mcK\left(|\Phi_0|^2,|\Phi_{\pi}|^2\right)=-3\ln\left\{F(|\Phi_\pi|^2)-F(|\Phi_0|^2)\right\}+\cdots,
\ee
where
\be
                    F(|\Phi|^2)=\int^{|\Phi|^2}dx\,x^{\frac{2}{3}\frac{k_0}{q_0-k_0}-1}\,(1-x)^{\frac{2}{3}}.
\ee
In view of this and of the fact that by fixing the $\Phi_i$ at values with $\Phi_0\neq\Phi_{\pi}$ we also fix the radion $T\sim\ln(\Phi_{\pi}/\Phi_0)$, one is naturally led to consider a supersymmetric version of the \emph{Goldberger-Wise} stabilization mechanism \cite{Goldberger:1999uk}, where now superpotentials $W_i(\Phi_i)$ localized at each brane ensure that the $\Phi_i$ are stabilized at different non-vanishing values. For example, one could take
\be
                   W_i(\Phi_i)=\ka_5^{-2}(\Phi_i-a_i)^2,
\ee   
with $a_0\neq a_\pi \neq 0$. In this model there is a supersymmetric Minkowski vacuum at $\Phi_i=a_i$, since
\be
                        D_i W_i(a_i)=W_i(a_i)=0,
\ee
and therefore
\be
                       V_F(a_i)=0.
\ee

Unfortunatly, this idea has a flaw. To understand that, it is usefull to rewrite the brane couplings in terms of the 5D superfields $h$ and $H$:
\be
                  \mcL_0= \delta(y)\,\int d^2\theta \,\phi^3\,W_0(\Phi_0)+\textup{h.c.}=\ka_5^2\,\delta(y)\,\int d^2\theta\, h^2\,W_0(H/h)+\textup{h.c.},
\ee 
\be
                   \mcL_\pi =\delta(y-\pi R)\,\int d^2\theta \,\phi^3\,W_\pi (\Phi_\pi )+\textup{h.c.} = \ka_5^2\,\delta(y-\pi R)\,\int d^2\theta\, h^2\,W_\pi (H/h)\,e^{-3k_0\pi T}+\textup{h.c.}
\ee
The problem is the presence of a non-local quantity, the "Wilson-line" $T\sim\oint dy\, \Sigma^0$, in the brane localized interaction $\mcL_\pi$. This is physically not sensible. The correct brane Lagrangian at $y=\pi R$ in fact must read\footnote{In addition to locality, the form of this F-term Lagrangian is determined by requiring the integrand composite chiral superfield in \eqref{eq:pi_lagran} to have the Weyl weight $w=3$. Since both $h$ and $H$ have $w=3/2$, it turns out that \eqref{eq:pi_lagran} is the most general brane Lagrangian satisfying this condition.}
\be\label{eq:pi_lagran}
                   \mcL_\pi = \ka_5^2\,\delta(y-\pi R)\,\int d^2\theta\, h^2\,W_\pi (H/h)+\textup{h.c.} =\delta(y-\pi R)\,\int d^2\theta \,\phi^3\,\,e^{3k_0\pi T}\,W_\pi \left(\Phi_0 \,e^{\frac{3}{2}(q_0-k_0)\pi T}\right)+\textup{h.c.},
\ee
and the effective 4D superpotential is thus
\be
                    W=W_0(\Phi_0)+e^{3k_0\pi T}\,W_\pi \left(\Phi_0 \,e^{\frac{3}{2}(q_0-k_0)\pi T}\right).
\ee

Supersymmetric ${\mathbb M}_4$ stable vacua can now be found by requiring $\partial_0 W=\partial_T W=W=0$. A simple model realizing this idea was proposed in \cite{Maru03} (see also \cite{Blechman04}), and consists of using linear brane superpotentials: $W_0=a\Phi_0$ and $W_\pi = -b\Phi_0 \,e^{\frac{3}{2}(q_0-k_0)\pi T}$. It is not difficult to find that there is a supersymmetric ${\mathbb M}_4$ vacuum at $\Phi_0=0$ and $T=\frac{2}{3\pi (q_0+k_0)}\ln(a/b)$. This vacuum is a minimum of the potential. In fact, it is a general feature of any supersymmetric Minkowski vacuum (in $\mcN=1$ SUGRA) that the mass-matrix is positive semi-definite\footnote{This has also been recently pointed out in \cite{Blanco-Pillado:2005fn}, in a study of racetrack superpotentials.}. One has in this case
\be
                  \partial_A\partial_{\bar B}V_F=M_P^4\,e^{\mcK}\mcK^{I{\bar J}}(\partial_A\partial_I W)(\partial_{\bar B}\partial_{\bar J} {\bar W}),
\ee 
showing that unless $\partial_A\partial_I W=0$ for $\forall I$ and some $A$, the matrix $\partial_A\partial_{\bar B}V_F$ is positive definite. In the particular case that we consider here, this matrix reads 
\be\label{eq:matrix}
\partial_A\partial_{\bar B}V_F=M_P^4\,e^{\mcK}\epsilon_{IA}\,\mcK^{I{\bar J}}\,\epsilon_{{\bar J}{\bar B}}\left|3(q_0+k_0)a/2\right|^2.
\ee

It is a nice feature of the model of \cite{Maru03} that there is \emph{no need for any fine-tuning} to obtain Minkowski susy vacua, unlike it is the case for racetrack models \cite{Blanco-Pillado:2005fn}. We would like, however, to express some concerns regarding the consistency of these models: Let us recall that both the compensator superfield $h$ and the physical superfield $H$ transform under the U(1) symmetry gauged by $V^0$ as
\be
                h\to\exp\left(\frac{3}{2}\epsilon(y)\,k_0\,\Lambda^0\right)h,\qquad H\to \exp\left(\frac{3}{2}\epsilon(y)\,q_0\,\Lambda^0\right)H,
\ee   
if $V^0\to V^0+\Lambda^0+\Lambda^{0+}$. Here, the chiral superfield $\Lambda^0(x,y)$ has odd orbifold parity. In particular, at the $y=0$ boundary we have 
\be
                h(x,0)\to\exp\left(\frac{3}{2}\,k_0\,\epsilon(0)\Lambda^0(x,0)\right)h(x,0),\qquad H(x,0)\to \exp\left(\frac{3}{2}\,q_0\,\epsilon(0)\Lambda^0(x,0)\right)H(x,0).
\ee
Thus, it is crucial to understand what the (value of the) (odd)$^2$ product $\epsilon(0)\Lambda^0(x,0)$ is. In fact, in case it does not vanish, to ensure the gauge invariance of the brane couplings ($\sim hH$) in the model of ref.\cite{Maru03} we must require the hypers to have opposite charges, that is $q_0+k_0=0$. Clearly, this has the consequence that the supersymmetric ${\mathbb M}_4$ minimum degenerates into a \emph{flat} potential, see e.g. eq.\eqref{eq:matrix}. We thus see that the consistency of these stabilization models depends crucially on the meaning of the product of ${\mathbb Z}_2$-odd functions at the orbifold boundaries. This is an interesting, in our view unsettled issue.

\locsection{Dimensional reduction of 5D heterotic M-theory}\label{sec:mtheory}

Having found in sec.\ref{sec:universal} the bulk superspace Lagrangians for the \emph{ungauged} 5D supergravity coupled to the universal hypermultiplet, we are now in a position to write down the 5D heterotic M-theory of \cite{Lukas:1998yy,Lukas:1998tt} in superfield language. The only modifications to the bulk Lagrangians \eqref{eq:univer_D} and \eqref{eq:univer_F} are due to the \emph{gauging} of an U(1) subgroup of the isometries of the universal hyperscalar manifold \cite{Fujita01}. Note that this U(1) isometry leaves $|H|^2-|h_1|^2$ invariant. To better account for these modifications let us introduce the following doublet
\be
               \mcH^{T}=\left(H,~h_1\right),
\ee    
and similar for $\mcH^c$. With this notation, eqs.\eqref{eq:univer_Dprim} and \eqref{eq:univer_Fprim} can be recasted as
\be
               \mcL_{D}= -3\int d^4\theta \,\mcN(V_y)^{\frac{1}{3}} \left[h_2^+e^{V_T}h_2  +h_2^{c+}e^{-V_T}h_2^c-\mcH^{\dagger}\sigma_3\,e^{-V_T}\mcH-\mcH^{c\dagger}\sigma_3\,e^{V_T}\mcH^c\right]^{\frac{2}{3}},
\ee
and
\be
               \mcL_{F}=-2\int d^2\theta \,\left\{h_2^c\partial_y h_2 -\mcH^{cT}\sigma_3\partial_y \mcH+\Sigma_T(h_2^c h_2+\mcH^{cT}\sigma_3 \mcH) \right\}+\textup{h.c.},
\ee
where $\sigma_3=\textup{diag}(1,-1)$.

The U(1) will be gauged by a combination ${\mathbb V}_S=\alpha_I{\mathbb V}^I$ of the vector multiplets with orbifold parities $(V,\Sigma)^I\sim(-,+)$. In the case of interest we can take \cite{Fujita01} the U(1) generator acting on $\mcH$ to be $\sigma_3+i\sigma_2$. Therefore, it will be usefull in the following to introduce the notation ${\bf V_S}=(\sigma_3+i\sigma_2)V_S$ and ${\bf \Sigma_S}=(\sigma_3+i\sigma_2)\Sigma_S$. The bulk Lagrangians for the 5D heterotic M-theory read
\be\label{eq:D_Mtheory} 
               \mcL_{D}= -3\int d^4\theta \,\mcN(V_y)^{\frac{1}{3}} \left[h_2^+e^{V_T}h_2  +h_2^{c+}e^{-V_T}h_2^c-\mcH^{\dagger}\sigma_3\,e^{-V_T-\epsilon(y){\bf V_S}}\mcH-\mcH^{c\dagger}\sigma_3\,e^{V_T+\epsilon(y){\bf V_S}}\mcH^c\right]^{\frac{2}{3}},
\ee
and
\be
               \mcL_{F}=-2\int d^2\theta \,\left\{h_2^c\partial_y h_2 -\mcH^{cT}\sigma_3\partial_y \mcH+\Sigma_T(h_2^c h_2+\mcH^{cT}\sigma_3 \mcH)+\epsilon(y)\mcH^{cT}\sigma_3{\bf \Sigma_S}\mcH \right\}+\textup{h.c.},
\ee
where $V_S=\alpha_I V^I$ and $\Sigma_S=\alpha_I\Sigma^I$. To make contact with the component (bulk) Lagrangians of \cite{Lukas:1998tt} one has to note that \cite{Fujita01}
\be\label{eq:intro_S}
               \left(\frac{H}{h_1}\right)_{\theta=0}=\frac{1-S}{1+S},\qquad \left(\frac{H^c}{h_2}\right)_{\theta=0}=\xi\frac{1+S}{S+S^+},
\ee
where $S,\xi$ are complex fields, and $S$ is related to the volume $V$ of the Calabi-Yau 3-fold by $S=V+|\xi|^2+i\sigma$. Here, $\sigma$ is the dual of the 3-form potential $C_{3}$ arising from the 11D bulk in M-theory. In addition one has also to solve the constraints arising from the variations of the multipliers $V_T$ and $\Sigma_T$.

\subsection{Effective 4D K\"ahler potential}

It is our goal now to obtain the effective low-energy 4D K\"ahler potential of this model. We will use the same strategy as in the previous section. We start by factorizing the y-dependence of $\mcH(x,y)$ and $\mcH^c(x,y)$ as 
\be\label{eq:rescaled_hyper} 
                \mcH=\exp\left(\int^y_0dy\epsilon(y){\bf\Sigma_S}\right)\Phi(x) =\left( 1+\int^y_0 dy \epsilon(y){\bf\Sigma_S}\right)\Phi(x),
\ee
\be
                \mcH^c=\left( 1-\int^y_0 dy \epsilon(y){\bf\Sigma_S}\right)\Phi^c(x),
\ee
where we used the fact that $(\sigma_3+i\sigma_2)^2=0$. Since $h_2^c$ and $\Phi^c$ are projected out by the orbifold twist they will not participate in the low-energy physics. For this reason we drop them in the following. Eq.\eqref{eq:D_Mtheory} reads now
\be
                \mcL_{D}= -3\int d^4\theta \,\mcN(V_y)^{\frac{1}{3}} \left[h_2^+e^{V_T}h_2-e^{-V_T}\Phi^{\dagger}\sigma_3\left(1+\int^y_0 dy \epsilon(y)(\Sigma_S+\Sigma_S^{+})(\sigma_3+i\sigma_2)\right)\Phi \right]^{\frac{2}{3}}.       
\ee
Integrating out the multiplier $V_T$ (see eq.\eqref{eq:multiplier}), writing $\Phi$ as 
\be
                   \Phi^T(x)=({\tilde h}_1 H_0,~{\tilde h}_1), \qquad \textup{and } H_0=(1-S_0)/(1+S_0)
\ee
and defining the 4D chiral compensator as $\phi^3\equiv \ka_5^2\,{\tilde h}_1h_2(1+H_0)$, we finally get
\be\label{eq:5Dmtheorytruncated}
                \mcL_{D}= -3\ka_5^{-\frac{4}{3}}\int d^4\theta \,\mcN(V_y)^{\frac{1}{3}} \phi^+\phi\left[S_0+S_0^+-2\int_0^ydy\epsilon(\Sigma_S+\Sigma_S^{+}) \right]^{\frac{1}{3}}. 
\ee
Note that the y-dependence enters this Lagrangian only through the vector scalars $\Sigma^I(x,y)$. There is again a simple case that can be handled in an exact way, the single modulus case with $\mcN=\ka_5 (M^0)^3$ and $\Sigma_S=\alpha_0\Sigma^0$. This corresponds to the special \emph{universal} case of \cite{Lukas:1998yy}. The corresponding 4D Lagrangian is thus
\be\begin{split}\label{eq:4Dmtheory}
                \mcL^{(4D)}=\oint dy \mcL_D=&-3\ka_5^{-1}\int d^4\theta \,\frac{3\phi^+\phi}{4\alpha_0}\left\{\left(S_0+S_0^+\right)^{\frac{4}{3}}\right.\\
		&\hspace{144pt}\left.-\left(S_0+S_0^+-2\pi\alpha_0(T+T^+)\right)^{\frac{4}{3}}\right\},
\end{split}\ee
where again
\be
                     T=\frac{1}{2\pi}\oint dy\, \Sigma^0.
\ee
We finaly obtain the effective K\"ahler potential
\be\label{eq:kaehler_mtheory}
                \mcK=-3\ln\frac{3}{4\alpha_0}\left[(S_0+{S_0}^{+})^{\frac{4}{3}}-(S_0+{S_0}^{+}-2\pi\alpha_0(T+T^{+}))^{\frac{4}{3}}\right].
\ee
It is noteworthy that this expression was previously only mentioned in \cite{Lalak:2001dv}. In fact, as far as we can say, all phenomenological and cosmological studies of 4D heterotic M-theory rely on the linearized approximation which corresponds to $S_0+{S_0}^{+}\gg 2\pi\alpha_0(T+T^{+})$. In this limit $\mcK(S_0,T)$ reduces to
\be\label{eq:limit1}
              \mcK\simeq-3\ln2\pi(T+T^+)-\ln(S_0+{S_0}^{+}),
\ee  
which is also the universal part of the low-energy effective K\"ahler potential of weakly coupled $E_8\times E_8$ string theory. This is not unexpected, as the condition $\textup{Re}(S_{0})\gg 2\pi\alpha_0 \textup{Re}(T)$ corresponds to the weak coupling limit of heterotic M-theory, where the size of the $S^1/Z_2$ orbifold is much smaller than the volume of the internal Calabi-Yau space.

Yet, from a phenomenological point of view, the most interesting case is the one with large warping. This corresponds to having an hierarchy in the sizes of the CY three-fold measured at the two boundaries of the orbifold. To gain a better understanding of this issue it is usefull to consider the chiral superfield $S$ introduced in eq.\eqref{eq:intro_S}, whose scalar component's real part $V$ measures the size of the CY 3-fold. It is not difficult to show, using \eqref{eq:rescaled_hyper}, that 
\be\label{eq:ydependS}
                  S(x,y)=S_0(x)-2\alpha_0\int_0^ydy\epsilon(y)\,\Sigma^0(x,y). 
\ee 
Let us thus introduce the notation $S_{\pi}(x)\equiv S(x,y=\pi R)=S_0-2\pi\alpha_0 T$, and rewrite the K\"ahler potential \eqref{eq:kaehler_mtheory} as\footnote{It happens often that instead of the moduli $S_0$ and $S_{\pi}$, one uses the moduli $T$ and $\tilde S$, where $T$ measures the size of the orbifold direction, while $\tilde S$ stands for the \emph{average} size of the CY 3-fold. In this case we have $\mcK=-3\ln\left[({\tilde S} +{\tilde S}^{+}+\pi\alpha_0(T+T^+))^{\frac{4}{3}}-({\tilde S} +{\tilde S}^{+}-\pi\alpha_0(T+T^+))^{\frac{4}{3}}\right]$.}
\be\label{eq:kaehler_newversion}
                  \mcK=-3\ln\frac{3}{4\alpha_0}\left[(S_0+{S_0}^{+})^{\frac{4}{3}}-(S_{\pi}+{S_{\pi}}^{+})^{\frac{4}{3}}\right].
\ee
It is clear now that a hierarchy between the sizes of the CY at the two branes implies $\textup{Re}(S_0)\gg\textup{Re}(S_{\pi})$, and in this limit the K\"ahler potential reads
\be\label{eq:limit2}
            \mcK\simeq-4\ln(S_0+S_0^+)+3\left(\frac{S_{\pi}+{S_{\pi}}^{+}}{S_0+S_0^+}\right)^{\frac{4}{3}}\simeq-4\ln(S_0+S_0^+)
\ee
It is remarkable how different the two limits \eqref{eq:limit1} and \eqref{eq:limit2} look like. And nevertheless, the "exact" K\"ahler potential displays some relevant properties that are \emph{independent} of the strength of the warping, namely we have (with $(i=0,\pi)$)
\be\label{eq:rel1}
            \mcK_i(S^i+S^{i+})=-4,
\ee
\be\label{eq:rel2}
            \mcK^{i{\bar j}}\mcK_{\bar j}=-(S^i+S^{i+}),
\ee
and therefore
\be\label{eq:almost_noscale}
            \mcK^{i{\bar j}}\mcK_i\mcK_{\bar j}=4.
\ee
We will see below that these properties extend also to the case with five-branes in the bulk. Eq.\eqref{eq:almost_noscale} implies, for example, that in the presence of a constant superpotential\footnote{A constant superpotential could be obtained e.g. as the remnant of a non-perturbative superpotential for frozen moduli.} $W_0$, the potential reads
\be
                V_F\sim \frac{|W_0|^2}{\left((s_0)^{\frac{4}{3}}-(s_{\pi})^{\frac{4}{3}}\right)^3}\simeq \frac{|W_0|^2}{(s_0)^{4}}\left(1+3\left(\frac{s_{\pi}}{s_0}\right)^{\frac{4}{3}}\right),
\ee
where $s_i(t)\equiv \textup{Re}(S_i)$. We see that a constant superpotential leads to a growth of the orbifold size and to the collapse of the size of the CY 3-fold at the hidden brane (at $y=\pi R$). It would be interesting to check if the addition of non-perturbative superpotentials, e.g. due to gaugino condensation at the orbifold branes, can lead to the stabilization of these moduli at dS vacua for $s_0\gg s_{\pi}$. We will pursue this issue in a separate paper \cite{usnew}.

Closing this section, we would like to comment on possible "dualities" existing in the absence of superpotentials. It is known that in the weak coupling limit defined by \eqref{eq:limit1}, the 4D action is invariant under certain SL(2,${\mathbb R}$) duality transformations: $T\to(aT-ib)/(icT+d)$ and similar for $S$, where the real parameters $a,b,c,d$ satisfy the identity $ad-bc=1$. These transformations leave the K\"ahler potential invariant up to a gauge K\"ahler transformation. It is clear that in the strongly coupled theory, with the K\"ahler potential given by eq.\eqref{eq:kaehler_mtheory} (or eq.\eqref{eq:kaehler_newversion}), these dualities do not persist. Moreover, we don't envisage a modification of the transformations leading to (generalized) dualities acting also in the strong coupling case. On the other hand, in the limit of very small CY size at the hidden brane, eq.\eqref{eq:limit2} displays again an SL(2,${\mathbb R}$) duality for $S_0$. We think it would be interesting to study this issue in more detail.

\subsection{Time-dependent solutions}\label{sec:timedep}

As a non-trivial \emph{test} of the validity of the K\"ahler potential \eqref{eq:kaehler_mtheory} we will present a family of 4D time-dependent (cosmological) solutions, which when uplifted to 5D will correspond to the recently \cite{Chen:2005jp} constructed exact time-dependent solutions of 5D heterotic M-theory.

There is a simple way of obtaining time-dependent 4D solutions from the supergravity Lagrangian \eqref{eq:4Dmtheory}. We will do this by sticking to the conformal frame $\phi=1+\theta^2 F_{\phi}$, as there is a set of \emph{time-dependent} solutions, for which in this frame the geometry is Minkowski. It is important to note that the contribution due to the non-trivial conformal factor $e^{-\frac{\mcK}{3}}$ to the Einstein equations of motion is quadratic in derivatives of $e^{-\frac{\mcK}{3}}$ \cite{carlos}
\be
                 \delta E_{\mu\nu}=-e^{\frac{\mcK}{3}}(\nabla_{\mu}\partial_{\nu}e^{-\frac{\mcK}{3}}-g_{\mu\nu}\Box e^{-\frac{\mcK}{3}}),
\ee
while the modifications to the equations of motion of the moduli $S_i$ are proportional to the Ricci scalar $R^{(4)}$. We see that configurations obtained by requiring that both the total moduli kinetic energy and the second time derivative of the conformal factor $e^{-\frac{\mcK}{3}}$ vanish, are solutions of the equations of motion in Minkowski space. 

The first requirement translates to
\be
                   \frac{|\partial_t S_0|^2}{(S_0+S_0^+)^{\frac{2}{3}}} = \frac{|\partial_t S_{\pi}|^2}{(S_{\pi}+S_{\pi}^+)^{\frac{2}{3}}},
\ee
which is solved by $s_i(t)=(a_i+b(t))^{\frac{3}{2}}$. The second requirement then implies that $b(t)=b\cdot t$, and we obtain
\be
                   s_i(t)=(a_i+b\,t)^{\frac{3}{2}}.
\ee

To understand from the 5D viewpoint what these solutions mean it is usefull to recall eq.\eqref{eq:ydependS}. This expression implies the following uplifting (oxidation) of the 4D time-dependent solutions to 5D
\be
                   ds_5^2=-e^{2\sigma}dt^2+(e^5_y)^2dy^2,
\ee
with
\be
                  e^5_y\sim S^{\frac{1}{3}}\sim e^{2\sigma}\sim(ay+bt)^{\half},
\ee
where $a=a_0/\pi R$ and we set $a_{\pi}=0$ by shifting $t$. As advertised, this is exactly the 5D time-dependent heterotic brane solution of ref.\cite{Chen:2005jp}. As pointed out by these authors, there is a curvature singularity at $ay+bt=0$, which for $a>0,b<0$, "invades" the $S^1/Z_2$ orbifold for $t>0$, corresponding to the collapse of the internal CY 3-fold. In the 4D effective theory this patology shows up in the form of a singular behaviour of the K\"ahler metric as $s_0(t)=(bt)^{\frac{3}{2}}$ approaches zero, even though the solutions are fully regular at $t=0$. Eventually, non-perturbative corrections to the low-energy heterotic M-theory will lead to the stabilization of the moduli away from these singular points in the K\"ahler manifold \cite{usnew}.

\subsection{Including bulk five-branes}\label{sec:fivebranes}

In addition to the boundary branes lying at the fix-points of the $S^1/Z_2$ orbifold, heterotic M-theory generally contains mobile five-branes which wrap holomorphic 2-cycles in the CY 3-fold and therefore appear as three-branes in the low-energy 5D bulk. As we will see now, it is straightforward, at least in the universal case ($h^{(1,1)}=1$), to obtain also the dependence of the effective K\"ahler potential on the moduli describing the fluctuations of the five-branes in the $S^1/Z_2$ direction. Here, we will consider first a single mobile brane, but it is straightforward to extend the result for a set-up with multiple five-branes, as we will see below.

Let us denote by $z$ the position of 5-brane along the y-direction. It is a crucial point of our analysis that a relevant part of the 5-brane interactions arise by a suitable modification of the odd couplings in eq.\eqref{eq:5Dmtheorytruncated}. To be precise, the effect of the inclusion of a 5-brane leads to the following replacement
\be\label{eq:chavetas}
                \alpha_0\epsilon(y)\to\alpha(y)=\left\{\begin{array}{ll} -\alpha_1 &, -\pi R<y<-z\\-\alpha_0 &, -z<y<0 \\ \alpha_0 &, 0<y<z\\ \alpha_1 &, z<y<\pi R\end{array}\right.
\ee   
Taking this into account in eq.\eqref{eq:5Dmtheorytruncated}, and integrating the modified expression over $y$, we obtain the following K\"ahler potential
\be\begin{split}\label{eq:alpha_y}
          \mcK=&-3\ln\frac{3}{4}\left[ \frac{1}{\alpha_0}\left(S_0+S_0^{+}\right)^{\frac{4}{3}}-\left(\frac{1}{\alpha_0}-\frac{1}{\alpha_1}\right)\left(S_0+S_0^{+}-2\alpha_0(Z+Z^+)\right)^{\frac{4}{3}}\right.\\
	       &\hspace{60pt}\left.-\frac{1}{\alpha_1}\left(S_0+S_0^{+}-2(\alpha_0-\alpha_1)(Z+Z^+)-2\pi\alpha_1 (T+T^{+}) \right)^{\frac{4}{3}}\right],         
\end{split}\ee 
where we introduced the superfield $Z\equiv \int_0^z\Sigma^0$, related to the position of the 5-brane. In the presence of additional 5-branes, additional terms appear in $\mcK$, which are proportional to powers of the CY volume measured at the different 5-branes. This K\"ahler potential should reduce to well-known expressions \cite{Derendinger:2000gy} in the limit of small warping, $\alpha_0,\alpha_1\to 0$, and it indeed does so:
\be
                   \mcK\simeq -3\ln2\pi(T+T^+)-\ln\left[S_0+S_0^+ +(\alpha_0-\alpha_1)\frac{(Z+Z^+)^2}{\pi(T+T^+)}\right].
\ee

As we already pointed out, also in the presence of 5-branes the derivatives of $\mcK$ fulfill a few relations which are coupling independent. With $S_5\equiv S_0-2\alpha_0 Z$ and $S_{\pi}\equiv S_0-2(\alpha_0-\alpha_1)Z-2\pi\alpha_1 T$, the K\"ahler potential is a function of $S_0,S_5,S_{\pi}$ (and conjugates), and we have again the relations \eqref{eq:rel1}, \eqref{eq:rel2} and \eqref{eq:almost_noscale}, where now the index $i$ runs over both the bulk and boundary branes ($i=0,5,\pi$). Clearly, these expressions extend also to multiple 5-branes in the bulk. Note that in the presence of a constant superpotential $W_0$ this leads to a growing orbifold size and a 5-brane sliding towards one of the boundaries (depending on the relative charges), until the size of the CY 3-fold at the hidden brane collapses.

But, also for $W_0=0$, we expect instabilities described by time-dependent solutions generalizing the ones presented in the previous section. Consider an arbitrary number of 5-branes in the $S^1/{\mathbb Z}_2$ bulk. The K\"ahler potential reads 
\be
                  \mcK=-3\ln\left(\sum_i\beta_i\left(S^i+S^{i+}\right)^{\frac{4}{3}}\right),
\ee
where the constants $\beta_i$ - one for each brane, including the boundary ones - satisfy the topological constraint $\sum_i\beta_i=0$. A simple set of solutions of the 4D equations of motion in the conformal frame is again
\be
                         s_i(t)=(a_i+b_i\, t)^{\frac{3}{2}},
\ee
where the $b_i$ must satisfy two constraints: $\sum \beta_i\,b_i=0$ and $\sum \beta_i\,b_i^2=0$. Clearly, $b_i=b$ is a solution. These oxidize to 5D solutions
\be
                   ds_5^2=-e^{2\sigma}dt^2+(e^5_y)^2dy^2,
\ee
with
\be
                  e^5_y\sim S^{\frac{1}{3}}\sim e^{2\sigma}\sim(a_0+bt-\gamma(y))^{\half},
\ee
where 
\be
                  \gamma(y)=\frac{2}{3}\ka_5^{-1}\int^y_0\,\alpha(y),
\ee
see eq.\eqref{eq:chavetas}. With this choice of coordinates, the positions $z_i$ of the bulk 5-branes are time-independent, being related to the integration constants $a_i$ and the brane charges $\alpha_i$ as
\be
                  z_i-z_{i-1}=\frac{3\ka_5}{2}\frac{a_{i-1}-a_i}{\alpha_{i-1}}.
\ee
However, the physical distance between the $i$-th and the $(i-1)$-th branes, given by the real part of 
\be
                 Z_i(t)-Z_{i-1}(t)=-\frac{1}{2\alpha_{i-1}}(S_i(t)-S_{i-1}(t)),
\ee
is time-dependent.

Clearly, the same kind of instability found in the heterotic M-theory setup without bulk branes by the authors of \cite{Chen:2005jp} is also present in the multi-brane system. For $b>0$, the curvature singularity appears first at the brane with lowest value of $\gamma(y)$ and then pervades the $S^{1}/{\mathbb Z}_2$ orbifold.

\subsection{Boundary couplings}\label{eq:boundary}

Finally, let us complete our analysis by presenting also the superspace description of the brane couplings of \cite{Lukas:1998yy} for the simplest case with a single $T$-modulus (i.e. Hodge number $h^{(1,1)}=1$). These are given by 
\be
                  \mcL_0=\frac{1}{4}\int d^2\theta S_0 \mcW\mcW+\textup{h.c.}+3\int d^4\theta\,\phi^+\phi\,(S_0+S_0^+)^{\frac{1}{3}}|C|^2+\int d^2\theta\,\phi^3\lambda_{ijk}C^iC^jC^k+\textup{h.c.},
\ee 
for the visible brane at $y=0$, and
\be
                  \mcL_{\pi}=\frac{1}{4}\int d^2\theta S_{\pi} \tilde{\mcW}{\tilde\mcW}+\textup{h.c.},
\ee
for the hidden brane at $y=\pi R$. Here $\mcW$ and $\tilde\mcW$ are the chiral superfield strengthes for the visible and hidden sector gauge theories, respectively, while the $C^i$ are charged matter chiral superfields at the visible brane. Recall that $S_{\pi}$, and therefore the holomorphic gauge coupling at the hidden brane, is given as 
\be
                  S_{\pi}=S_0-2(\alpha_0-\alpha_1)Z-2\pi\alpha_1 T,
\ee
in case we have a single 5-brane, in agreement with \cite{Derendinger:2000gy}. The corrections proportional to the moduli $Z$ and $T$, which from the 5D supergravity point of view are due to the warping of the geometry, are interpreted on the heterotic string side as one-loop \emph{threshold} corrections (see e.g. \cite{Nilles:1997vk}).  

Let us see now how the inclusion of these brane couplings modifies the 4D low-energy action. It is rather suggestive to combine $\mcL_D$ of eq.\eqref{eq:5Dmtheorytruncated} with the above kinetic term for the matter chiral superfields $C^i$:
\be
                \mcL_D'= -3\ka_5^{-1}\int d^4\theta \,\left(\Sigma^0+\Sigma^{0+}-\partial_yV^0-\ka_5\delta(y)|C|^2\right) \phi^+\phi\left[S(x,y)+S^{+}(x,y)+2\alpha_0\epsilon(y)V^0\right]^{\frac{1}{3}},
\ee
where we re-introduced the \emph{odd} vector superfield $V^0$, omitted in  \eqref{eq:5Dmtheorytruncated}. The "dilaton" superfield $S$ is defined in \eqref{eq:ydependS}. We now introduce a new real superfield ${\tilde V}^0$ as
\be
          {\tilde V}^0=V^0+\ka_5|C|^2\,\epsilon(y)\frac{\pi R-|y|}{2\pi R},      
\ee
and rewrite $\mcL_D'$ as
\be
                 \mcL_D'= -3\ka_5^{-1}\int d^4\theta \,\left(\Sigma^0+\Sigma^{0+}-\frac{\ka_5}{2\pi R}|C|^2\right) \phi^+\phi\left[S(x,y)+S^{+}(x,y)-\frac{\alpha_0\ka_5}{\pi R}(\pi R-|y|)|C|^2\right]^{\frac{1}{3}}.   
\ee
Note that now we \emph{dropped} the odd real superfield ${\tilde V}^0$. The reason for this is simple: ${\tilde V}^0$ vanishes at the boundaries and hence does not contribute to the low-energy action (cf. discussion before eq.\eqref{eq:4.9}). In fact, it is the vector superfield $V^0$ that acquires a VEV, which due to the brane-localised coupling \cite{Falkowski:2005fm} is non-vanishing at the boundary, and it is this VEV that now appears in $\mcL_D'$. It is noteworthy that the effect of the brane-localised term has become a bulk effect. 

To obtain the effective K\"ahler potential we can proceed now as we did before, since 
$\mcL_D'$ is a total derivative. Recalling that $S=S_0-\alpha_0\int_0^ydy\epsilon(y)\,\Sigma^0(x,y)$ and integrating over the $S^1/Z_2$ direction, we finally obtain 
\be
               \mcK=-3\ln\frac{3}{4\alpha_0}\left[\left(S_0+{S_0}^{+}-\alpha_0\ka_5|C|^2\right)^{\frac{4}{3}}-(S_0+{S_0}^{+}-2\pi\alpha_0(T+T^{+}))^{\frac{4}{3}}\right].
\ee
Again, it is interesting to compare this expression to known ones in the limit of small warping (small $\alpha_0$). One gets
\be
               \mcK\simeq -\ln(S_0+S_0^+)-3\ln(T+T^+)+\left(\frac{3}{2\pi(T+T^+)}+\frac{\alpha_0}{2(S_0+S_0^+)}\right)\,\ka_5|C|^2+\frac{\alpha_0\pi(T+T^+)}{S_0+S_0^+}.
\ee
Notice that we omited other terms which despite being of order $\mcO(\alpha_0)$ are of higher order in $|C|^2$. Apart from the last term on the r.h.s., which is new, this expression agrees with the one obtained in \cite{Lukas:1997fg}. This result can be easily extended to include multiple five-branes in the bulk. For example, for a single five-brane we have
\be\begin{split}
                        \mcK=&-3\ln\frac{3}{4}\left[\alpha_0^{-1}\left(S_0+{S_0}^{+}-\alpha_0\ka_5|C|^2\right)^{\frac{4}{3}}-\alpha_0^{-1}\left(S_5+{S_5}^{+}-\alpha_0\ka_5\frac{\pi R-z}{\pi R}|C|^2\right)^{\frac{4}{3}}\right.\\
			&\hspace{120pt}\left.+\alpha_1^{-1}\left(S_5+{S_5}^{+}-\alpha_1\ka_5\frac{\pi R-z}{\pi R}|C|^2\right)^{\frac{4}{3}}-\alpha_1^{-1}(S_\pi+{S_\pi}^{+})^{\frac{4}{3}}\right].
\end{split}\ee

Finally, the superpotential entails two terms,
\be
                 W=\lambda_{ijk}C^iC^jC^k+W_{np},
\ee
where the non-perturbative piece arises from gaugino condensation at the hidden brane (see \cite{Lukas:1997rb}), which reads
\be                
        	W_{gc}=g_\pi \frac{S_\pi}{S_0}\exp\left(-a_\pi S_\pi \right),
\ee
where $g_\pi$ and $a_\pi$ are superfield independent constants, and from open membrane instantons stretching between the branes \cite{Moore:2000fs,Lima:2001jc}. It is expected that a combination of these effects will lead to moduli stabilization in a phenomenologically interesting ground state (see e.g. \cite{Becker:2004gw}). We will study this issue in a separate publication \cite{usnew}.

\locsection{Conclusions}

A first important step in this paper was the integration of the real radion supermultiplet ${\mathbb W}_y$ which brought us much closer to the usual superfield description of global 5D SUSY with a radion chiral multiplet. It was an easy walk to reproduce the effective 5D superaction of \cite{Marti:2001iw} and to go beyond that. It is also no problem to enlarge the compensator sector and to obtain an universal hypermultiplet. This then also allows to discuss 5D heterotic M-theory in superfields and to obtain several nice new results. 

Within this formalism, the transition from 5D to 4D can be achieved in a very elegant and transparent way because of the underlying superconformal structure. The treatment of warping, improving the previous studies of \cite{us04a}, was a central issue here. We provided a carefull analysis that goes beyond the linear (small warping) approximation, obtaining several exact results. These include the effective K\"ahler potential of heterotic M-theory for $h^{(1,1)}=1$ with an arbitrary number of bulk five-branes, and the determination of time-dependent (moving branes) solutions generalizing the ones of ref.\cite{Chen:2005jp}. It is surely a challenge to extend these successes to even more complex scalar manifolds. 

Alltogether, our new approach should allow to deal with multi-moduli models in a very elegant way, both in the question of stabilization \cite{usnew} and - perhaps even more interesting - in the discussion of the dynamical behaviour of branes and of other moduli fields important for inflationary cosmology.

\paragraph{Note added:} As we were concluding the redaction of this paper, we received ref.\cite{Arroja:2006zz} where the time-dependent solutions of 4D heterotic M-theory presented in section \ref{sec:timedep}, which oxidize to the 5D solutions of \cite{Chen:2005jp}, were independently found albeit using a different approach. Their study does not include the \emph{new} time-dependent solutions of heterotic M-theory with bulk 5-branes, that we obtained in section \ref{sec:fivebranes}.

\vs{0.5cm}

\hs{-0.7cm}{\bf Acknowledgments}

\vs{0.2cm} 
\hs{-0.7cm}We thank C. Herdeiro, A. Hebecker and M. Olechowski for discussions and interesting comments. Z.T. aknowledges the Centro de F\'isica do Porto for hospitality while visiting Porto University.
The research of F.P.C. is supported by Funda\c c\~ ao para a 
Ci\^ encia e a Tecnologia (grant   SFRH/ BPD/20667/2004).

\appendix

\locsection{Generic instability of tree-level vacua}\label{app:instability}

Without resorting to constant brane superpotentials it is not possible, at tree-level, to find stable vacua, i.e. minima of the vector moduli potential. We give here a detailed proof of this fact. To be more precise, we will show that the potential
\be
                 V_F =-M_P^4{\bar g}_I{\tilde\mcN}^{IJ}{\bar g}_{J}-\frac{M_P^4}{2\tilde\mcN}({\bar g}_I(\Sigma^I+\Sigma^{I+}))^2,
\ee
has no minima at \emph{finite} values of the moduli $\Sigma^I$. We first note that 
\be
                  \frac{\partial V_F}{\partial\Sigma^I}(\Sigma^I+\Sigma^{I+})=-V_F,
\ee
where we used that 
\be
                     \frac{\partial{\tilde\mcN}^{IJ}}{\partial\Sigma^L}=-{\tilde\mcN}^{IM}{\tilde\mcN}_{KLM}{\tilde\mcN}^{KJ},
\ee 
as well as ${\tilde\mcN}_{I}=\frac{1}{2}{\tilde\mcN}_{IJ}(\Sigma+\Sigma^+)^J$ and ${\tilde\mcN}=\frac{1}{3}{\tilde\mcN}_{I}(\Sigma+\Sigma^+)^I$. So, clearly if $\Sigma_0^I$ is a minimum of $V_F$, the potential must vanish at that point. Using this fact, in a second step we find that there is a \emph{vector} $v^{ J}\equiv{\bar g}_I{\tilde\mcN}^{IJ}(\Sigma_0+\Sigma_0^+)$, such that
\be
               v^J\frac{\partial^2 V_F}{\partial\Sigma_0^J\partial{\bar\Sigma}_0^{\bar I}}v^I=-\frac{M_P^4}{27{\tilde\mcN}^3}({\bar g}_I(\Sigma_0+\Sigma_0^{+})^I)^4\leq 0.
\ee
This completes the proof that at tree-level, for ${\bar g}_I\neq 0$, there are no minima at finite (non-zero) values of the moduli.

\end{document}